\documentclass[]{aastex63}
\usepackage{graphicx}
\usepackage{advdate}
\usepackage{amsmath}
\usepackage{appendix}
\usepackage{xcolor}

\newcommand{\asec}{\hbox to 1pt{}\rlap{$^{\prime\prime}$}.\hbox to 2pt{}}
\newcommand{\amin}{\hbox to 1pt{}\rlap{$^{\prime}$}.\hbox to 1pt{}}
\newcommand{\adeg}{\hbox to 1pt{}\rlap{$^{\circ}$}.\hbox to 2pt{}}
\newcommand{\mbf}{\rm }
\newcommand{\tbf}{\rm }
\newcommand{\rbf}{\rm }

\accepted{for publication in {\it The Astrophysical Journal}}
\shortauthors{Postman, Lauer, Parker et al.}
\shorttitle{The Cosmic Optical Background}

\begin{document}

\title{New Synoptic Observations of the Cosmic Optical Background with New Horizons }

\author{Marc Postman}
\affil{Space Telescope Science Institute,\footnote{Operated by AURA, Inc., for the National Aeronautics and Space Administration}
3700 San Martin Drive, Baltimore, MD 21218}

\author{Tod R. Lauer}
\affil{U.S. National Science Foundation National Optical Infrared Astronomy Research
Laboratory,\footnote{The NSF NOIRLab is operated by AURA, Inc.
under cooperative agreement with NSF.}
P.O. Box 26732, Tucson, AZ 85726}

\author{Joel W. Parker}
\affil{Department of Space Studies, Southwest Research Institute, 1050 Walnut St., Suite 300, Boulder, CO 80302}

\author{John R. Spencer}
\affil{Department of Space Studies, Southwest Research Institute, 1050 Walnut St., Suite 300, Boulder, CO 80302}

\author{Harold A. Weaver}
\affil{The Johns Hopkins University Applied Physics Laboratory,
Laurel, MD 20723-6099}

\author{J. Michael Shull}
\affil{Department of Astrophysical \& Planetary Sciences, CASA, University of Colorado, Boulder, CO 80309}
\affil{Department of Physics \& Astronomy, University of North Carolina, Chapel Hill, NC 27599}

\author{S. Alan Stern}
\affil{Space Science and Engineering Division, Southwest Research Institute, 1050 Walnut St., Suite 300, Boulder, CO 80302}

\author{Pontus Brandt}
\affil{The Johns Hopkins University Applied Physics Laboratory,
Laurel, MD 20723-6099}

\author{Steven J. Conard}
\affil{The Johns Hopkins University Applied Physics Laboratory,
Laurel, MD 20723-6099}

\author{G. Randall Gladstone}
\affil{Southwest Research Institute, San Antonio, TX 78238}
\affil{University of Texas at San Antonio, San Antonio, TX 78249}

\author{Carey  M. Lisse}
\affil{The Johns Hopkins University Applied Physics Laboratory,
Laurel, MD 20723-6099}

\author{Simon B. Porter}
\affil{Department of Space Studies, Southwest Research Institute, 1050 Walnut St., Suite 300, Boulder, CO 80302}

\author{Kelsi N. Singer}
\affil{Department of Space Studies, Southwest Research Institute, 1050 Walnut St., Suite 300, Boulder, CO 80302}

\author{Anne. J. Verbiscer}
\affil{University of Virginia, Charlottesville, VA 22904}

\correspondingauthor{Marc Postman, postman@stsci.edu}

\begin{abstract}

We obtained New Horizons LORRI images to measure the cosmic optical background (COB) intensity integrated  over $0.4\lesssim\lambda\lesssim0.9{~\rm\mu m}.$ The survey comprises 16 high Galactic-latitude fields selected to minimize scattered diffuse Galactic light (DGL) from the Milky Way galaxy, as well as scattered light from bright stars. This work supersedes an earlier analysis based on observations of one of the present fields. Isolating the COB contribution to the raw total sky levels measured in the fields requires subtracting the remaining scattered light from bright stars and galaxies, intensity from faint stars within the fields fainter than the photometric detection-limit, and the DGL foreground. DGL is estimated from Planck HFI $350 {~\rm\mu m}$ and $550 {~\rm\mu m}$ intensities, using a new self-calibrated indicator based on the 16 fields augmented with eight additional DGL calibration fields obtained as part of the survey. The survey yields a highly significant detection {\mbf ($6.8\sigma$)} of the COB at {\mbf ${\rm 11.16\pm 1.65~(1.47~sys,~0.75~ran) ~nW ~m^{-2} ~sr^{-1}}$} at the LORRI pivot wavelength of 0.608 $\mu$m. The estimated integrated intensity from background galaxies, ${\rm 8.17\pm 1.18  ~nW ~m^{-2} ~sr^{-1}},$ can account for the great majority of this signal. The rest of the COB signal, {\mbf ${\rm 2.99\pm2.03~ (1.75~sys,~1.03~ran) ~nW ~m^{-2} ~sr^{-1}},$} is formally classified as anomalous intensity but is not significantly different from zero. The simplest interpretation is that the COB is completely due to galaxies.

\end{abstract}


\section{Light From a Dark Universe}

At the dawn of time, the Universe was a sea of light.  But as it expanded, it cooled, dimmed, and matter came to the fore.  Nearly 14 billion years after the Big Bang, space is now cold and dark.  While our horizon encompasses almost a trillion galaxies that have formed over that time, they are terribly faint, and we need our most powerful telescopes to tally their presence directly.  But their stars and accreting  black holes contribute in whole or part to a background of visible light that pervades the Universe. Paradoxically, it can be detected with a small telescope simply stationed at a suitable vantage point.

This cosmic optical background (COB) testifies to all processes that have generated light over the history of the Universe.  Is the COB intensity as expected from our census of faint galaxies, or does the Universe contain additional sources of light not yet recognized? We have attempted to detect the COB with the Long-Range Reconnaissance Imager \citep[LORRI;][]{lorri, lorri2} onboard NASA's New Horizons spacecraft as it explores the outer limits of the Kuiper Belt, bound for the depths of the interstellar space beyond. At 57 AU from the Sun, it is the most remote camera ever deployed. And it looks out into the darkest skies ever seen.

At New Horizons' (NH) location, the sky is essentially free of the zodiacal light (ZL) foreground, which is sunlight scattered by interplanetary dust. ZL strongly dominates the sky brightness in the inner solar system and has bedeviled all attempts to measure the COB from Earth-space. \citet{zemcov} however, recognized that the LORRI camera on NH could be useful for COB observations, and from the limited archival observations available at the time, recovered an upper limit to the COB intensity somewhat lower than the low-significance COB measurements based on HST, CIBER, and other observations.

\citet[hereafter NH21]{nh21} measured the COB from deep LORRI archival images obtained after the NH Pluto encounter. Based on seven fields at 42 to 45 au from the Sun, we measured the COB intensity to be in the range ${\rm 15.9\pm 4.2\ (1.8~stat., 3.7~sys.) ~nW ~m^{-2} ~sr^{-1}}$ at the LORRI pivot wavelength of 0.608 $\mu$m.{\footnote{\tbf We presented two COB levels in NH21, based on two different DGL estimators \citep{zemcov,brandt}.  We only quote the one using the \citet{zemcov} estimator, which is more compatible with the new DGL estimators derived here.} When the estimated integrated light of galaxies (IGL) fainter than the LORRI photometric detection-limit was subtracted from this intensity, a component of unknown origin in the range ${\rm 8.8\pm4.9\ (1.8 ~stat., 4.5 ~sys.) ~nW ~m^{-2} ~sr^{-1}}$ remained. 

The NH21 analysis also showed that, while the strong zodiacal light foreground was eliminated by the great distance of NH from the Sun, corrections for other foreground light sources were still required. The strongest of these was diffuse Galactic light (DGL), which is Milky Way starlight scattered into the line of sight by interstellar dust. We also had to correct for scattered starlight (SSL) from bright stars outside the LORRI field of view. DGL and SSL vary strongly over the sky, however, which means that fields can be targeted that greatly minimize the contributions of both foregrounds, compared to what is available from the random COB sampling provided by archival images observed for other purposes.

The present work is based on a program of new LORRI images obtained explicitly to minimize foreground contributions to the COB intensity observations. The program also incorporates improved understanding of how to use LORRI for low light-level imaging, as well as calibration observations obtained to improve understanding of the DGL and SSL foregrounds. 

In 2021 we tested our observational strategy by imaging one of the 16 COB fields that define the present program.  That test field, designated as NHTF01, verified our ability to select for greatly reduced DGL and SSL foregrounds.  Its analysis yielded a highly significant detection of the COB at ${\rm 16.37\pm 1.47  ~nW ~m^{-2} ~sr^{-1}}$ \citep[hereafter NH22]{nh22}.  Intriguingly, the estimated intensity due to all background galaxies (IGL) accounted for only half of this signal, implying that the COB also includes an intensity component of unknown origin at ${\rm 8.06\pm1.92 ~nW ~m^{-2} ~sr^{-1}}.$  This conclusion was supported by an independent analysis of archival LORRI observations by \citet{symons}, which recovered a total COB intensity of ${\rm 21.98\pm 1.23~(ran) \pm 1.36~(sys)  ~nW ~m^{-2} ~sr^{-1}},$ a result that implied the existence of an even larger anomalous component.

With the complete COB survey, however, we have reworked the estimation of the DGL foreground, finding its contribution to be stronger than the DGL estimates used in NH21 and NH22.  This now reduces the total COB intensity estimate, and more than halves both the estimated anomalous component and its significance. While our observations can accommodate a modest COB anomaly relative to the amplitude of the IGL, we cannot falsify the simpler hypothesis that the COB is due entirely to the known population of galaxies.

{\tbf Although the present paper reaches a qualitatively different conclusion about the COB than did NH21 and NH22, it does heavily rely on the analysis developed in those two earlier papers.  We summarize the interrelationship of all three papers in fine detail in Table \ref{tab:program}.  The first column shows the sample selection, image analysis, and foreground components treatments developed and presented in NH21, with the sections of that paper identified where the given item was discussed in detail.  The second column shows where NH22 augments or revises the analysis details in NH21.  The final column shows where the present paper augmented or revised our previous works.}

\begin{deluxetable}{lll}
\tabletypesize{\scriptsize}
\tablecolumns{3}
\tablewidth{0pt}
\tablecaption{\tbf The New Horizons Cosmic Optical Background Program}
\tablehead{
 & & \\
NH21 \citep{nh21} & NH22 \citep{nh22} & This work
}
\startdata
 & & \\
Sample Selection: $\S2$ SEA $\geq95^\circ$ & & \\
\phantom{Sample Selection: }$\S2$ Delay 150s & & \\
\phantom{Sample Selection: }$\S2$ Gal lat $|b| \geq50^\circ$ &$\S2.1$ Gal lat $|b|\geq40^\circ$& \\
\phantom{Sample Selection: }$\S2$ Exp time 30s&$\S2.2$ Exp time 65s& \\
\phantom{Sample Selection: }$\S2$ Number of fields: 7&$\S2.1$ Number of fields: 1&$\S2.1$ Number of fields: 16 $+$ 8 DCAL\\
 &$\S2.1$ Minimize DGL, SSL for field selection & \\
\hline
 & & \\
 Single image detection limit: $\S4.2$ $V=19.1$&$\S2.4$ $V=19.9$& \\
 \hline
 & &  \\
 Environment: $\S2.1$ Spacecraft Shadow & \\
\phantom{Environment:} $\S2.1.1$ Thruster Exhaust & \\
 &$\S{\rm A.1}$ Cherenkov (RTG $\gamma$-rays) \\
 &$\S{\rm A.2}$ Fluorescence (RTG $\gamma$-rays) \\
 &$\S{\rm A.3}$ Cherenkov (RTG scattered $\gamma$-rays) \\
 &$\S{\rm A.4}$ Cherenkov (RTG Neutrons)&  \\
 &$\S{\rm A.5}$ Cherenkov (Cosmic Rays)& \\
 \hline
 & & \\
 Image Analysis: $\S2.1$ Power-on Fade & & \\
 \phantom{Image Analysis:} $\S3.1.1$ Bias level & & \\
 \phantom{Image Analysis:} $\S3.1.2$ Dark current & & \\
 \phantom{Image Analysis:} $\S3.1.3$ Bias structure & & \\
 \phantom{Image Analysis:} $\S3.1.3$ Jail bars& & \\
 \phantom{Image Analysis:} $\S3.1.4$ Charge smear& & \\
 \phantom{Image Analysis:} $\S3.1.4$ Flat fielding & &  \\
 \phantom{Image Analysis:} $\S3.2.1$ Object masking & & \\
 \phantom{Image Analysis:} $\S3.2.2$ Sky measurement& & \\
 &$\S2.3$ A/D correction &$\S2.2$ No A/D correction needed\\
 & &$\S2.5$ Background decay \\
 \hline
 & & \\
 Foregrounds: $\S4.3$ Scattered starlight & & \\
 \phantom{Foregrounds:} $\S4.1$ Integrated starlight & & \\
 \phantom{Foregrounds:} $\S4.2$ Integrated galaxy light & & \\
 \phantom{Foregrounds:} $\S4.3$ SGL = 0.07 $\rm nW~m^{-2}~sr^{-1}$& &$\S4.3$ SGL = 0.10 $\rm nW~m^{-2}~sr^{-1}$\\
  & $\S3.7$ Bright galaxies (PS1) &$\S3.1$ Bright galaxies (DLS) \\
 \phantom{Foregrounds:} &$\S3.4$ Two photon continuum (2PC) & $\S3.4$ No 2PC correction needed\\
  & & $\S3.4$ H$\alpha$ emission (WHAM)\\
  \hline
 & & \\
 DGL: $\S4.4$ FIR band $100 {~\rm\mu m}$ (IRIS)& &$\S4.3$ FIR $350 {~\rm\mu m}~+$ $550 {~\rm\mu m}$ (Planck) \\
\phantom{DGL:} $\S4.4$ FIR aperture $0\adeg2$ radius circle& &$\S4.2$ FIR aperture LORRI FOV \\
\phantom{DGL:} $\S4.4.1$ Residual ZL correction ({\it error}) & &$\S4$ No residual ZL correction needed \\
\phantom{DGL:} $\S4.4$ Estimator: Zemcov; Brandt \& Draine &$\S3.2$ Zemcov only&$\S4.3$ NCOB $+$ DCAL self calibration\\
\enddata
\tablecomments{The first column lists the major components of the COB measurement presented in the NH21 (cross-referenced by the section in the paper in which it was discussed). The second column shows where the second paper, NH22, introduced new major components (those with no previous NH21 entry) or revised the analysis in NH21 (a previous NH21 entry was given).  The final column show how the present paper revised or augmented the analysis present in NH22 and/or NH21. We marked the the NH21 residual ZL correction entry with {\it ``error"} as its use in NH21 and NH22 was incorrect. {\bf A blank column entry to the right of a filled column entry means no change was made from the previous work.}}
\end{deluxetable}\label{tab:program}

\begin{figure}[hbtp]
\centering
\includegraphics[keepaspectratio,width=7.0 in]{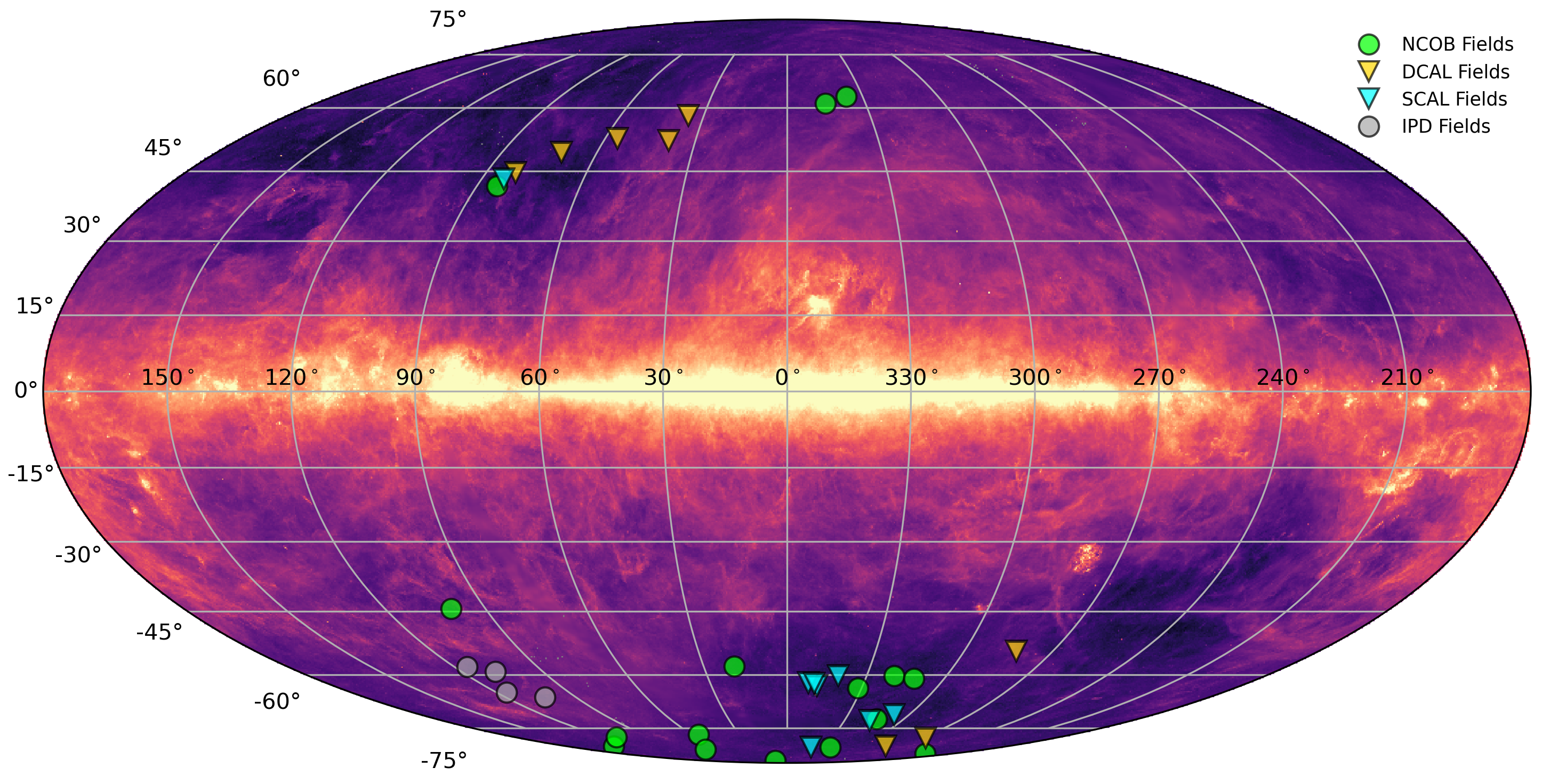}
\caption{The locations of the NCOB, DCAL, {\mbf SCAL, and IPD} fields are shown on the IRIS full-sky $100 {~\rm\mu m}$ map in Galactic coordinates.}
\label{fig:full_sky}
\end{figure}
\pagebreak
\section{The COB Survey}

\subsection{The Survey Design and Field Selection}\label{sec:design}

A truly cosmological optical background should present no more structure than would be expected given the known projected correlation function of large scale structure marked by galaxies over the age of the Universe.  More plainly, we expect the field-to-field variation of observed COB intensity to be more or less isotropic, once known instrumental effects and Galactic foregrounds have been corrected for. An ideal COB survey would attempt to uniformly sample the full sky on all angular scales. However, foregrounds associated with the Milky Way, or even the solar system, are markedly anisotropic over large angular scales. Thus any realistic survey will have additional Galactic and ecliptic coordinate constraints to minimize those foregrounds. Within such constraints, however, we are free to identify survey areas that specifically minimize both the DGL and SSL foregrounds, which are the two dominant foreground signatures for our observing platform.

In the case of the present survey, several considerations limit us to observing in somewhat restricted regions close to the Galactic poles. The strongest restriction is that needed to avoid scattered sunlight entering the LORRI aperture.  This issue is discussed in detail in NH21; briefly, we require the aperture to be fully in the shadow of the NH spacecraft.  This requirement means selecting fields with solar elongation angle (SEA) $>95^\circ.$ While SEA~$>90^\circ$ would be sufficient to keep direct sunlight out of the LORRI aperture, the spacecraft bulkhead in which the aperture is positioned also supports other instruments that could potentially scatter sunlight into LORRI; SEA~$>95^\circ$ ensures that these are also shaded by the spacecraft \citep[NH22]{lorri2}.

The trajectory of NH out of the solar system was completely specified by its primary mission of obtaining the first exploration of Pluto \citep{pluto} and the Kupier belt object Arrokoth \citep{arrokoth}. At the time of the mission, Pluto as seen from Earth was projected against the bulge of the Milky Way. This means that the ``anti-solar" hemisphere accessible to NH is roughly centered on the heart of our galaxy.  Requiring Galactic latitude $|b|>40^\circ$ to avoid dense stellar foregrounds and strong dust absorption thus eliminates a significant fraction of this hemisphere.  Lastly, we restrict ecliptic latitude to $|\beta|>15^\circ.$ While NH is not directly affected by zodiacal light, the FIR intensities that we use to select COB fields are provided by maps made in Earth-space, and thus may incur larger errors near the ecliptic \citep{akari2011, skysurf, korngut2022}.

The combination of these three constraints (SEA $>95^\circ$, $|b|>40^\circ$, and $|\beta|>15^\circ$) leaves 4239 $\rm deg^2$ of sky available. We randomly selected 60,000 positions within this area, and for each one estimated the DGL contribution using the IRIS 100$~\mu$m all-sky data and the amount of scattered starlight entering the LORRI field of view. The details of how we derived these preliminary foreground light estimates are discussed in NH22 (see $\S2.1$ of that work). We then ranked each position by the sum of these two foreground contributions and selected the 500 positions with the smallest sums for further review. From these 500 positions, we reduced the sample to 15 fields, to ensure we could perform the desired observations within the available spacecraft fuel constraints. We also ensured that there was coverage in both Galactic hemispheres and that the pointings spanned a broad range in Galactic longitude. The final coordinates of each field were adjusted by up to $0\adeg7$ to minimize the presence of bright galaxies or galaxy clusters within the LORRI field of view\footnote{As a consequence of this small tweaking of field position, one field - NCOB12 - falls slightly below the $|\beta| = 15^{\circ}$ limit.}. We also used the NH22 test field (NHTF01) as part of the survey since, by definition, it meets all the above requirements. This brings the total number of COB survey fields to 16. With the exception of NHTF01, we denote the COB science fields with the prefix NCOB.

We selected an additional set of eight fields to perform an improved self-calibration of the relation of FIR intensity to optical DGL. The FIR-DGL relation calibration fields are denoted with the prefix DCAL. The DCAL fields were explicitly selected to cover fields with progressively higher $100~\mu$m surface brightness, up to a limit of $\sim 3~{\rm MJy~sr^{-1}}$. {\tbf This limit was selected to avoid dust optical depths large enough that non-linear behavior between the FIR intensity and scattered light amplitude might start to come into play.} All DCAL fields were selected to have SSL levels in a narrow range ($6 < \rm SSL < 9~{\rm nW~m^{-2}~sr^{-1}}$) that was similar to the NCOB science fields. 

We also selected eight fields to help verify the scattered starlight estimates. These fields are denoted with the prefix SCAL. The SCAL fields were chosen to be closer to brighter stars than we would otherwise permit for the science observations to test the reliability of the scattered light estimates. All SCAL fields were selected to have DGL levels in a narrow range ($3 < \rm DGL < 6~{\rm nW~m^{-2}~sr^{-1}}$) that is similar to that in the NCOB science fields. 

Lastly, we selected four fields at low ecliptic latitude ($| \beta | \leq 6^{\circ}$) solely to verify the lack of significant brightness from interplanetary dust at the large heliocentric distance of the NH spacecraft at the time of our observations. These four fields are denoted with the prefix IPDF.

The coordinates, observation dates, and spacecraft heliocentric distance for the 16 COB fields, 16 calibration fields, and 4 IPD fields are listed in Table~\ref{tab:coords}. The MET (mission elapsed time) image identifiers of the first image of each field are also given. We show the field distribution on the sky with respect to the IRIS $100~\mu$m map in Figure~\ref{fig:full_sky}. The $30' \times 30'$ regions of sky from the DESI/DECam Legacy Survey \citep[hereafter DLS]{dls} centered on our 16 COB science fields are shown in Figure~\ref{fig:dls_fields}. The intentional lack of bright optical sources in any of the fields is clearly demonstrated.  Table~\ref{tab:irpars} provides the Galactic extinction $E(B-V)$, HI column density, H$\alpha$ emission, dust temperature, and the FIR and cosmic IR background (CIB) intensities for each field. The CIB-subtracted FIR intensities are used to estimate the DGL foregrounds for each field. While we list the $E(B-V)$ \citep{SF11} values for each field in Table~\ref{tab:irpars} these extinction indicators are not used in any of the analyses in this work. \cite{shull_pan24} have demonstrated that these $E(B-V)$ values are, on average, 12\% lower than those derived from Planck far-infrared observations. All errors listed are 1-$\sigma$ values.

\begin{deluxetable}{lrrrrrccc}
\tabletypesize{\scriptsize}
\tablecolumns{9}
\tablewidth{0pt}
\tablecaption{Survey Field Centers and Observations}
\tablehead{
\multicolumn{9}{c}{~~}\\
\multicolumn{1}{c}{Field} &
\multicolumn{2}{c}{R.A.~~(J2000)~~Dec.} &
\multicolumn{1}{c}{Galactic} &
\multicolumn{1}{c}{Galactic} &
\multicolumn{1}{c}{Ecliptic} &
\multicolumn{1}{c}{$r_h$} &
\multicolumn{1}{c}{Date} &
\multicolumn{1}{c}{MET} \\
\multicolumn{1}{c}{ID} &
\multicolumn{1}{c}{(deg)} &
\multicolumn{1}{c}{(deg)} &
\multicolumn{1}{c}{Longitude} &
\multicolumn{1}{c}{Latitude} &
\multicolumn{1}{c}{Latitude} &
\multicolumn{1}{c}{(au)} &
\multicolumn{1}{c}{(UT)} &
\multicolumn{1}{c}{(s)}
}
\startdata
NHTF01&  0.0756&$-21.5451$& 55.794&$-77.094$&$-19.720$&51.3&2021-09-24&0494832182\\
NCOB01&358.4334&$-54.9137$&319.728&$-60.293$&$-48.117$&57.0&2023-09-14&0556982941\\
NCOB02&  5.3540&$-55.6590$&311.662&$-60.958$&$-51.123$&56.9&2023-08-31&0555788941\\
NCOB03&353.7867&$-49.1893$&331.516&$-63.490$&$-41.781$&56.8&2023-08-22&0554999401\\
NCOB04&  8.0987&$-44.4906$&314.112&$-72.222$&$-43.074$&56.8&2023-08-21&0554933281\\
NCOB05& 10.7611&$-27.3461$& 25.794&$-88.122$&$-29.172$&56.8&2023-08-29&0555599521\\
NCOB06&  9.4350&$-34.7328$&323.189&$-81.850$&$-35.192$&57.0&2023-09-13&0556884241\\
NCOB07& 19.0398&$-26.6161$&209.027&$-84.464$&$-31.807$&56.8&2023-08-28&0555537901\\
NCOB08&336.2651&$-30.0473$& 18.925&$-57.870$&$-18.712$&56.9&2023-08-30&0555723121\\
NCOB09&  6.7398&$-22.1689$& 73.341&$-82.551$&$-22.919$&56.9&2023-08-30&0555661141\\
NCOB10& 15.7115&$-18.8994$&141.203&$-81.364$&$-23.521$&56.8&2023-08-27&0555476401\\
NCOB11& 10.6266&$-15.2837$&112.541&$-77.975$&$-18.216$&56.8&2023-08-27&0555414901\\
NCOB12&207.4692&$  3.9649$&336.539&$ 62.959$&$ 14.270$&56.8&2023-08-17&0554613781\\
NCOB13&211.9528&$  4.6995$&345.372&$ 61.111$&$ 16.557$&56.8&2023-08-17&0554548681\\
NCOB14&356.2651&$ 15.5111$&100.270&$-44.420$&$ 15.684$&56.8&2023-08-20&0554838541\\
NCOB15&247.9273&$ 55.2059$& 84.133&$ 41.702$&$ 74.534$&56.7&2023-08-13&0554210281\\
DCAL01& 17.0453&$-34.9907$&279.471&$-81.361$&$-38.437$&56.8&2023-08-20&0554871721\\
DCAL02& 21.3854&$-36.2819$&266.706&$-78.332$&$-41.288$&57.0&2023-09-12&0556822921\\
DCAL03&239.1696&$ 44.8625$& 71.217&$ 49.235$&$ 62.786$&56.7&2023-08-13&0554274001\\
DCAL04&243.8895&$ 52.4461$& 81.173&$ 44.640$&$ 70.921$&56.7&2023-08-13&0554242381\\
DCAL05&236.0924&$ 34.9450$& 55.856&$ 52.438$&$ 52.762$&56.7&2023-08-14&0554305561\\
DCAL06& 36.3653&$-58.9600$&282.325&$-54.245$&$-65.192$&57.0&2023-09-13&0556916881\\
DCAL07&228.2392&$ 24.1691$& 35.663&$ 58.073$&$ 40.267$&56.7&2023-08-15&0554368501\\
DCAL08&235.2073&$ 24.5594$& 38.839&$ 51.978$&$ 42.723$&56.7&2023-08-14&0554337061\\
SCAL01&  6.8703&$-33.6610$&339.789&$-81.676$&$-33.240$&57.0&2023-09-12&0556790881\\
SCAL02&344.2125&$-42.5989$&351.834&$-62.077$&$-32.811$&57.0&2023-09-12&0556783201\\
SCAL03&346.2247&$-43.7754$&348.135&$-62.860$&$-34.496$&57.0&2023-09-12&0556786801\\
SCAL04&  7.1526&$-43.8832$&316.818&$-72.624$&$-42.197$&57.0&2023-09-11&0556781041\\
SCAL05&245.6015&$ 54.3330$& 83.387&$ 43.205$&$ 73.058$&56.7&2023-08-12&0554149081\\
SCAL06&345.0136&$-43.5540$&349.450&$-62.190$&$-33.892$&57.0&2023-09-12&0556785001\\
SCAL07& 10.7746&$-46.1545$&307.349&$-70.902$&$-45.497$&57.0&2023-09-11&0556770781\\
SCAL08&345.8313&$-48.0810$&340.769&$-60.336$&$-38.144$&57.0&2023-09-12&0556788661\\
IPDF01 &$ 4.9830 $&$ 2.6940 $&$ 107.415 $&$ -59.225 $&$ 0.493 $&56.8&2023-08-19& 0554765221 \\
IPDF02 &$ 8.8178 $&$ 4.6502 $&$ 115.318 $&$ -57.982 $&$ 0.780 $&56.8&2023-08-19& 0554767681 \\
IPDF03 &$ 9.6595 $&$ -1.8759 $&$ 115.469 $&$ -64.562 $&$ -5.552 $&56.8&2023-08-19& 0554762701 \\
IPDF04 &$ 4.9272 $&$ -4.3385 $&$ 103.170 $&$ -65.985 $&$ -5.939 $&56.8&2023-08-19& 0554760181 \\
\enddata
\tablecomments{All coordinates are in degrees. The $r_h$ parameter is the distance of the spacecraft from the Sun at the time of the observation. MET is the mission elapsed time in seconds of the first image in each field sequence. NCOB fields are the primary fields for measuring the COB intensity. NHTF01 is the test of the NCOB field selection and observational strategy published earlier in NH22. DCAL and SCAL fields are for DGL and scattered starlight calibration, respectively.  IPD fields are the low-ecliptic latitude fields taken to verify the lack of zodi-emission at these large heliocentric distances. }
\end{deluxetable}\label{tab:coords}

\begin{figure}[hbtp]
\centering
\includegraphics[keepaspectratio,width=7.0 in]{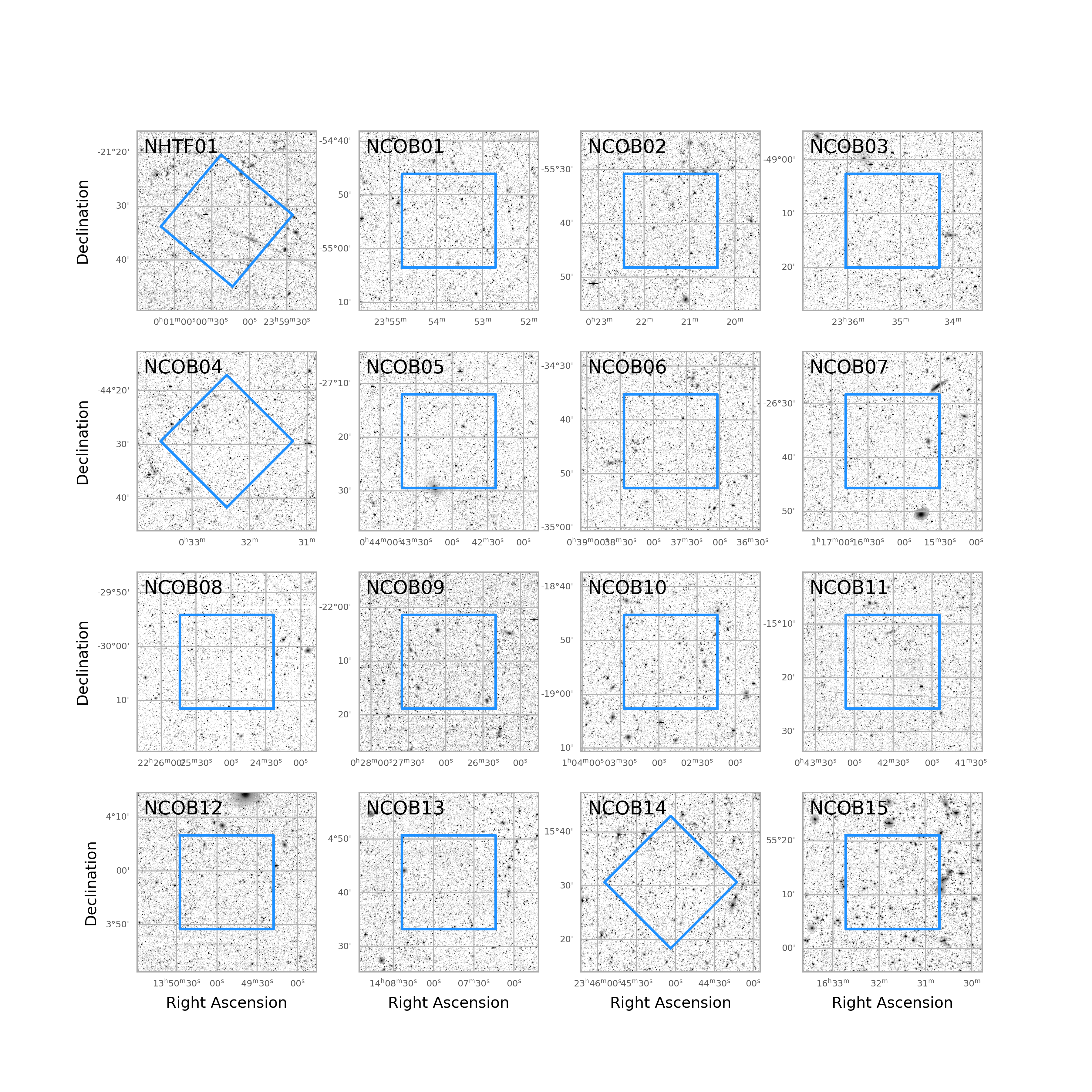}
\caption{The positions of the NCOB fields are shown with respect to images from the Deep Legacy Survey \citep{dls}. The DLS image cutouts above show a 30 arcmin field of view centered on the LORRI position. The blue boxes show the LORRI CCD orientation relative to J2000 equatorial coordinates. The streak in the NHTF01 DLS image is an artifact. Some fields were rolled in position angle to minimize spacecraft maneuvers.}
\label{fig:dls_fields}
\end{figure}

\subsection{Images of the Fields}

Images of the survey fields were obtained with LORRI.  With the exception of those for NHTF01, the observations were obtained over August and September of 2023 when the spacecraft was nearly 57 au from the Sun. For each COB survey field (those designated with the NCOB prefix), 16 65s LORRI exposures were obtained, while eight 65s exposures were obtained of the DGL calibration (DCAL) and scattered starlight calibration (SCAL) fields. The imaging sequences for each of the four IPD fields consisted of 16 65s LORRI exposures. To avoid the LORRI ``background fade" anomaly associated with the activation of the camera (NH21), no images were obtained earlier than four minutes after LORRI was powered on. 

LORRI is an unfiltered (white light) $1032\times1024$ pixel CCD imager mounted on a 20.9 cm aperture Cassegrain reflector. The active imaging area is $1024\times1024$ pixels, with the final eight columns being covered by a dark shield.  The last four shielded columns are used to measure the combined dark current and electronic bias level. For deep observations, such as those used for COB measurements, the camera is operated with $4\times4$ pixel binning, producing raw images in $257\times256$ pixel format, including a single bias/dark column. The pixel-scale in this mode is $4\asec08,$ which provides a $17\amin4$ field. LORRI's sensitivity extends from the blue ($0.4~\mu{\rm m}$) to NIR ($0.9~\mu{\rm m}$) and is defined by the CCD response and telescope optics.  The pivot wavelength is 0.608$~\mu$m.  The gain is $19.4e^-$ per 1 DN, and the read-noise is $24e^-.$ In $4\times4$ mode, the photometric zeropoint is $18.88\pm0.01$ AB magnitudes corresponding to a 1 DN/s exposure level \citep{lorri2}.

The suitability of LORRI for COB observations is discussed at length in NH21 and NH22. In short, we have explored the spacecraft environment for potential extraneous contributions to the background sky level, as well as similar effects in the LORRI camera. As shown in NH21, the spacecraft shadow is adequate for preventing both direct and indirect solar illumination of the LORRI aperture.  Analysis in NH21 also shows that the thrusters that control the spacecraft attitude do not generate particulates around the spacecraft that scatter sunlight.  Lastly, $\gamma$-rays emitted by the spacecraft's radioisotope thermoelectric generator (RTG) or cosmic rays do not generate significant intensities of optical photons through the Cerenkov or fluorescent mechanisms (see NH22).

 As LORRI is operated, the electronic bias level and average dark-current level are measured from the over-scan column as a combined electronic background level. In NH21 we presented the analysis of a novel calibration sequence of exposures that allowed for direct isolation of the dark-level. This demonstrated that the dark current was as expected, based on pre-launch calibration tests.  In NH21 we further demonstrated that bias images obtained with LORRI produced a null background. In NH22 we discussed our discovery of a low-level error in the LORRI analog/digital conversion electronics, {\tbf which caused the bias level to be in error by 0.02 DN. Subsequent analysis shows that it affects other signal measurements at the same level, thus it has no significant net effect and is ignored in the present analysis.}
\pagebreak
\subsection{Image Reduction}

The sky levels in the images are only slightly greater than 1 DN. Accurate recovery of the sky intensity thus requires attention to subtle effects that become important at this level.  As detailed in NH21 and NH22, we use a custom image reduction pipeline to optimize accurate recovery of the faint sky signal. This includes estimating the bias level by fitting a gaussian to the peak of the DN histogram of the bias column. We also include a special step to correct for the ``jail bar" pattern, in which the bias level of the even-numbered columns in the CCD is offset by either $+0.5$ or $-0.5$ DN from that of the odd-numbered columns (the sign of the correction varies randomly between LORRI power-on cycles). Lastly, we exclude bright cosmic ray hits and negative amplifier ``under-shoot" artifacts associated with over-exposed stars from the LORRI charge-smear corrections, as they are not smeared.

\subsection{Measuring the Sky Level}

The procedures for measuring the sky levels are discussed extensively in NH21. In brief, we measure the sky for each individual exposure by first masking out foreground stars, galaxies, hot pixels, and cosmic ray events, and then fitting a gaussian to the peak of the intensity histogram of the remaining unmasked pixels. The histogram fitting algorithm is designed to take into account fine scale structure in the distribution of pixel intensity values that results from the image calibration operations applied to the initially integer raw pixel values.  We emphasize that the final sky value for a given field is the the average of the individual sky levels measured for the 16 or eight images obtained of the field, as opposed to the single sky value measured from a stack of all the images.  As noted in \citet{lorri2} and NH21, LORRI exhibits a slowly varying pattern of row-wise low-amplitude ($<1$~DN) streaks in its bias level. This pattern is treated as a random noise source, which is captured in the dispersion of sky values measured for any field.

\begin{figure}[hbtp]
\centering
\includegraphics[keepaspectratio,width=6.0 in]{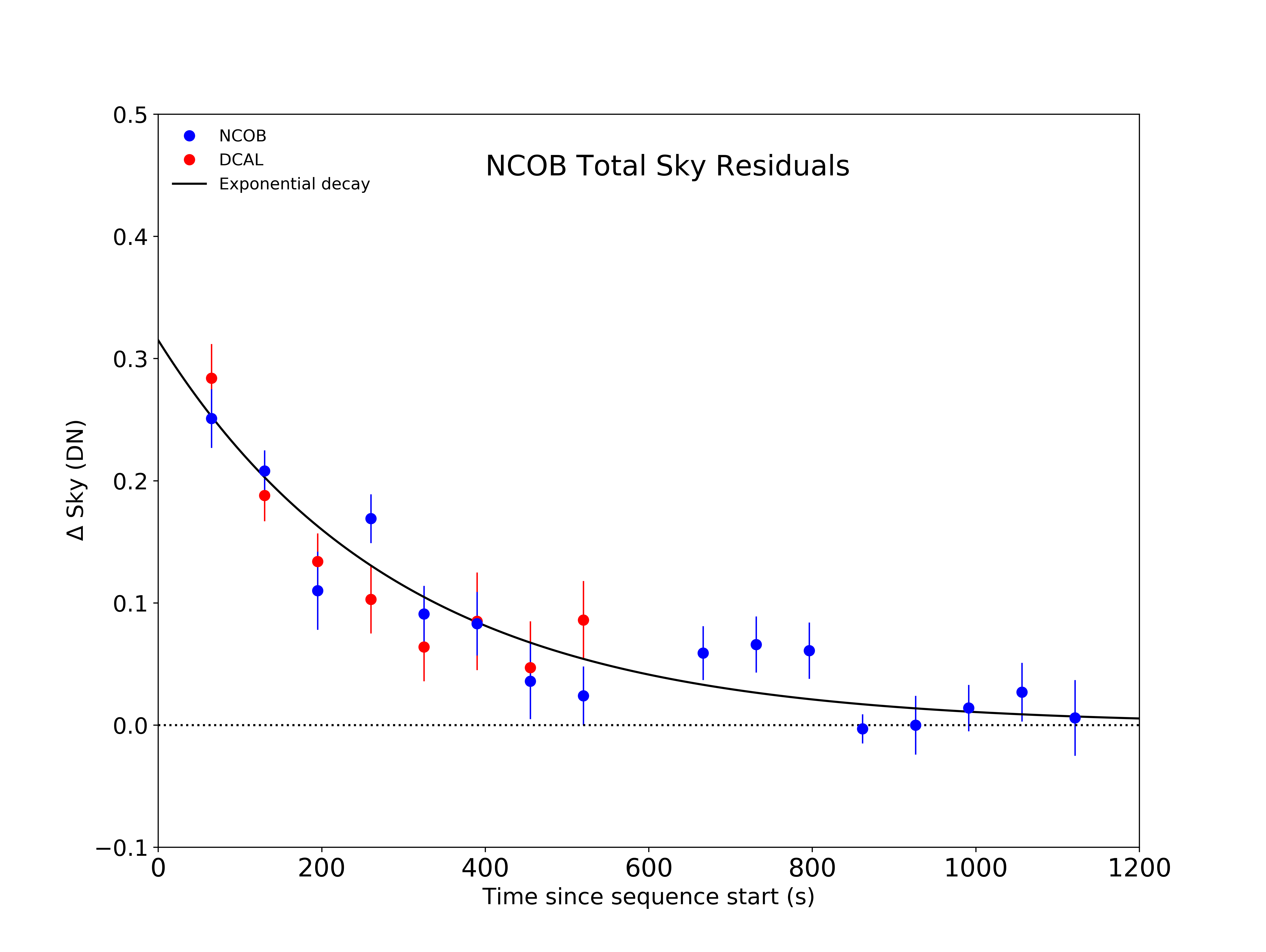}
\caption{The plot shows the average residuals in the total sky level for each image obtained for a given field, as a function of time since the sequence start {\tbf under the assumption that the sky level is constant over the sequence of images for any given field.} The 15 NCOB fields (blue) each comprise 16 images, and the eight DCAL (red) fields comprise eight images. The curve is an exponential decay model fitted to the average NCOB residuals. The reference level for the DCAL residuals is taken as the average of the model over the first eight NCOB images.}
\label{fig:drift}
\end{figure}

\subsection{Background Decay}\label{sec:decay}

In NH21 we discovered that images taken shortly after LORRI was powered on had elevated background levels, which appeared to decay away during the initial four minutes of operation. The cause of this background effect is unknown. The NH22 test COB images were therefore obtained only after this ``cool-down" interval had elapsed following activation of the instrument. As noted in NH22, the sky levels in the 16 images obtained of the test field appeared to be constant over the sequence, validating this solution.  We thus used the same procedure to obtain the present data.

Examining the sky levels measured for each image in our present richer data set, however, we have found that the background decay still continues even after the four minute delay, albeit at a low level.  Comparison of the amplitude of the decaying background between the NCOB and DCAL exposures, which covers roughly a factor of two in total sky level, shows that the background excess was not tied to the exposure level. It is thus modeled as an additive effect.

All NCOB exposures comprise the same sequence of eight 65s images taken in rapid succession, followed by an 80s pause to adjust the spacecraft pointing, followed by the final eight images, again taken in rapid succession.  Subtracting the mean total sky level from the complete set of 16 images for any field showed {\tbf that} the first images {\tbf had} generally positive residuals {\tbf compared to the average level over the sequence}, with the final images having slightly negative residuals.  Figure \ref{fig:drift} shows the average residual trend for all the NCOB fields as a function of time.

An exponential decay model appears to be an excellent description of the behavior of the residuals with time.  We fitted the trace as:
\begin{equation}
\Delta_{sky}(t)=ae^{-t/\tau}+b. \label{eqn:skydecay}
\end{equation}
where $\Delta_{sky}(t)$ is the average sky residual at any position on the NCOB exposure sequence, $t$ is the time since the start of the sequence, and $b$ is a constant background, which accounts for the fact that the initial mean sky for any field will include the background excess. For the NCOB sequence, a least-squares fit recovers $a=0.315$ DN,
$\tau=295$ s, and $b=-0.075$ DN, which corresponds to ${\rm 1.72 ~nW ~m^{-2} ~sr^{-1}}$ in intensity units. The background term in essence is the correction needed for the initial total sky levels to account for the presence of the decaying background. In practice, we use the model to apply a correction to each image in the sequence, which removes a source of variance in determination of the final average level, thus reducing the random error in the total sky measures.  We have also now applied this correction to NHTF01, with the caveat that its images were taken over a longer interval, thus reducing its net correction to $-0.061$ DN, or ${\rm 1.39 ~nW ~m^{-2} ~sr^{-1}}.$

{\tbf The final  decay model fitted to the uncorrected residuals is shown in Figure \ref{fig:drift}.} We also include the sky residuals from the DCAL images in the figure.  These fields have only eight images, and the shorter duration of the exposure sequence following the sequence start means that they are affected more strongly by the decaying background. The exponential model also provides the correction for these residuals, and as can be seen, are in excellent agreement with the NCOB images taken at the same time lag. {\tbf Use of the decay correction does add a systematic uncertainty of $0.16 {\rm~nW~m^{-2}~sr^{-1}}$ to the the sky level for any field, which is included in the total error budget.}

\begin{deluxetable}{lCCCCCCCC}
\tabletypesize{\scriptsize}
\tablecolumns{9}
\tablewidth{0pt}
\tablecaption{Field Properties for Dust, Gas, and FIR Emission}
\tablehead{
\multicolumn{9}{c}{~}\\
\multicolumn{1}{c}{Field} &
\multicolumn{1}{c}{$E(B-V)$} &
\multicolumn{1}{c}{N(HI)} &
\multicolumn{1}{c}{$\rm H\alpha$} &
\multicolumn{1}{c}{$\rm T_{Dust}$} &
\multicolumn{1}{c}{I(350 $\mu$m)} &
\multicolumn{1}{c}{CIB (350 $\mu$m)} &
\multicolumn{1}{c}{I(550 $\mu$m)} &
\multicolumn{1}{c}{CIB (550 $\mu$m)} \\
\multicolumn{1}{c}{ID} &
\multicolumn{1}{c}{(mag)} &
\multicolumn{1}{c}{($\rm 10^{20}~cm^{-2}$)} &
\multicolumn{1}{c}{(R)} &
\multicolumn{1}{c}{(deg K)} &
\multicolumn{1}{c}{(MJy sr$^{-1}$)} &
\multicolumn{1}{c}{(MJy sr$^{-1}$)} &
\multicolumn{1}{c}{(MJy sr$^{-1}$)} &
\multicolumn{1}{c}{(MJy sr$^{-1}$)} 
}
\startdata
  NHTF01 &  0.014 &   1.613 &   0.517 $\pm$  0.035 &  21.73 $\pm$  0.04 &  0.918 $\pm$ 0.046 &  0.521 $\pm$ 0.035 &  0.450 $\pm$ 0.023 &  0.343 $\pm$ 0.010 \\
  NCOB01 &  0.007 &   1.040 &   0.705 $\pm$  0.041 &  17.31 $\pm$  0.03 &  0.799 $\pm$ 0.040 &  0.451 $\pm$ 0.034 &  0.390 $\pm$ 0.020 &  0.303 $\pm$ 0.009 \\
  NCOB02 &  0.008 &   1.259 &   0.783 $\pm$  0.042 &  17.79 $\pm$  0.04 &  1.053 $\pm$ 0.053 &  0.632 $\pm$ 0.034 &  0.503 $\pm$ 0.025 &  0.398 $\pm$ 0.009 \\
  NCOB03 &  0.008 &   1.083 &   0.867 $\pm$  0.045 &  17.91 $\pm$  0.03 &  0.934 $\pm$ 0.047 &  0.565 $\pm$ 0.034 &  0.459 $\pm$ 0.023 &  0.370 $\pm$ 0.009 \\
  NCOB04 &  0.005 &   3.337 &   1.017 $\pm$  0.047 &  16.93 $\pm$  0.03 &  0.824 $\pm$ 0.041 &  0.557 $\pm$ 0.034 &  0.426 $\pm$ 0.021 &  0.355 $\pm$ 0.009 \\
  NCOB05 &  0.008 &   1.222 &   0.428 $\pm$  0.039 &  20.75 $\pm$  0.02 &  0.975 $\pm$ 0.049 &  0.677 $\pm$ 0.034 &  0.499 $\pm$ 0.025 &  0.429 $\pm$ 0.009 \\
  NCOB06 &  0.010 &   1.908 &   0.646 $\pm$  0.042 &  20.20 $\pm$  0.02 &  0.978 $\pm$ 0.049 &  0.585 $\pm$ 0.034 &  0.471 $\pm$ 0.024 &  0.376 $\pm$ 0.009 \\
  NCOB07 &  0.011 &   1.348 &   0.324 $\pm$  0.035 &  20.70 $\pm$  0.06 &  1.001 $\pm$ 0.050 &  0.574 $\pm$ 0.034 &  0.485 $\pm$ 0.024 &  0.374 $\pm$ 0.009 \\
  NCOB08 &  0.013 &   1.045 &   0.693 $\pm$  0.039 &  21.53 $\pm$  0.07 &  0.913 $\pm$ 0.046 &  0.578 $\pm$ 0.034 &  0.456 $\pm$ 0.023 &  0.370 $\pm$ 0.009 \\
  NCOB09 &  0.013 &   1.440 &   0.549 $\pm$  0.035 &  22.82 $\pm$  0.03 &  1.014 $\pm$ 0.051 &  0.647 $\pm$ 0.035 &  0.500 $\pm$ 0.025 &  0.406 $\pm$ 0.010 \\
  NCOB10 &  0.012 &   1.143 &   0.314 $\pm$  0.034 &  23.01 $\pm$  0.15 &  0.912 $\pm$ 0.046 &  0.628 $\pm$ 0.034 &  0.481 $\pm$ 0.024 &  0.399 $\pm$ 0.009 \\
  NCOB11 &  0.012 &   1.396 &   0.390 $\pm$  0.034 &  24.03 $\pm$  0.01 &  0.853 $\pm$ 0.043 &  0.534 $\pm$ 0.034 &  0.420 $\pm$ 0.021 &  0.350 $\pm$ 0.009 \\
  NCOB12 &  0.019 &   1.833 &   0.408 $\pm$  0.032 &  20.59 $\pm$  0.02 &  1.243 $\pm$ 0.062 &  0.535 $\pm$ 0.034 &  0.567 $\pm$ 0.028 &  0.349 $\pm$ 0.009 \\
  NCOB13 &  0.020 &   1.916 &   0.452 $\pm$  0.033 &  21.14 $\pm$  0.01 &  1.204 $\pm$ 0.060 &  0.581 $\pm$ 0.034 &  0.543 $\pm$ 0.027 &  0.371 $\pm$ 0.009 \\
  NCOB14 &  0.017 &   2.213 &   0.973 $\pm$  0.033 &  18.57 $\pm$  0.08 &  1.357 $\pm$ 0.068 &  0.524 $\pm$ 0.034 &  0.591 $\pm$ 0.030 &  0.335 $\pm$ 0.009 \\
  NCOB15 &  0.005 &   1.502 &   0.459 $\pm$  0.036 &  17.60 $\pm$  0.06 &  0.890 $\pm$ 0.045 &  0.554 $\pm$ 0.034 &  0.460 $\pm$ 0.023 &  0.357 $\pm$ 0.009 \\
  DCAL01 &  0.016 &   1.770 &   0.484 $\pm$  0.043 &  20.91 $\pm$  0.01 &  1.194 $\pm$ 0.060 &  0.668 $\pm$ 0.034 &  0.584 $\pm$ 0.029 &  0.419 $\pm$ 0.009 \\
  DCAL02 &  0.019 &   1.947 &   0.975 $\pm$  0.057 &  20.90 $\pm$  0.02 &  1.195 $\pm$ 0.060 &  0.552 $\pm$ 0.034 &  0.551 $\pm$ 0.028 &  0.368 $\pm$ 0.009 \\
  DCAL03 &  0.017 &   1.466 &   0.740 $\pm$  0.035 &  19.50 $\pm$  0.07 &  1.138 $\pm$ 0.057 &  0.530 $\pm$ 0.035 &  0.519 $\pm$ 0.026 &  0.351 $\pm$ 0.010 \\
  DCAL04 &  0.019 &   1.751 &   0.633 $\pm$  0.034 &  19.03 $\pm$  0.02 &  1.361 $\pm$ 0.068 &  0.569 $\pm$ 0.034 &  0.607 $\pm$ 0.030 &  0.369 $\pm$ 0.009 \\
  DCAL05 &  0.024 &   2.021 &   0.698 $\pm$  0.034 &  20.73 $\pm$  0.02 &  1.387 $\pm$ 0.069 &  0.615 $\pm$ 0.034 &  0.596 $\pm$ 0.030 &  0.390 $\pm$ 0.009 \\
  DCAL06 &  0.028 &   2.951 &   0.886 $\pm$  0.053 &  19.43 $\pm$  0.02 &  1.808 $\pm$ 0.090 &  0.545 $\pm$ 0.034 &  0.770 $\pm$ 0.039 &  0.364 $\pm$ 0.009 \\
  DCAL07 &  0.037 &   3.557 &   0.812 $\pm$  0.039 &  18.78 $\pm$  0.01 &  2.044 $\pm$ 0.102 &  0.488 $\pm$ 0.034 &  0.832 $\pm$ 0.042 &  0.327 $\pm$ 0.009 \\
  DCAL08 &  0.038 &   3.646 &   0.852 $\pm$  0.033 &  19.78 $\pm$  0.01 &  2.047 $\pm$ 0.102 &  0.545 $\pm$ 0.034 &  0.843 $\pm$ 0.042 &  0.355 $\pm$ 0.009 \\
  SCAL01 &  0.010 &   1.197 &   0.604 $\pm$  0.044 &  19.68 $\pm$  0.07 &  0.911 $\pm$ 0.046 &  0.606 $\pm$ 0.035 &  0.460 $\pm$ 0.023 &  0.386 $\pm$ 0.010 \\
  SCAL02 &  0.008 &   1.006 &   0.919 $\pm$  0.046 &  19.07 $\pm$  0.05 &  0.935 $\pm$ 0.047 &  0.581 $\pm$ 0.034 &  0.472 $\pm$ 0.024 &  0.378 $\pm$ 0.009 \\
  SCAL03 &  0.010 &   0.882 &   0.762 $\pm$  0.043 &  17.94 $\pm$  0.01 &  0.901 $\pm$ 0.045 &  0.556 $\pm$ 0.034 &  0.446 $\pm$ 0.022 &  0.355 $\pm$ 0.009 \\
  SCAL04 &  0.009 &   1.805 &   0.723 $\pm$  0.043 &  16.88 $\pm$  0.01 &  0.953 $\pm$ 0.048 &  0.629 $\pm$ 0.034 &  0.470 $\pm$ 0.024 &  0.393 $\pm$ 0.009 \\
  SCAL05 &  0.007 &   1.133 &   0.290 $\pm$  0.035 &  18.05 $\pm$  0.08 &  0.903 $\pm$ 0.045 &  0.541 $\pm$ 0.034 &  0.451 $\pm$ 0.023 &  0.350 $\pm$ 0.009 \\
  SCAL06 &  0.008 &   0.946 &   0.868 $\pm$  0.043 &  18.28 $\pm$  0.03 &  0.889 $\pm$ 0.044 &  0.550 $\pm$ 0.034 &  0.451 $\pm$ 0.023 &  0.360 $\pm$ 0.009 \\
  SCAL07 &  0.010 &   2.384 &   0.688 $^{*}$\phantom{~~0.000} &  17.83 $\pm$  0.03 &  1.015 $\pm$ 0.051 &  0.634 $\pm$ 0.034 &  0.495 $\pm$ 0.025 &  0.397 $\pm$ 0.009 \\
  SCAL08 &  0.009 &   0.997 &   0.825 $\pm$  0.044 &  18.22 $\pm$  0.06 &  0.968 $\pm$ 0.048 &  0.618 $\pm$ 0.034 &  0.470 $\pm$ 0.024 &  0.395 $\pm$ 0.009 \\
  IPDF01 &  0.020 &   3.038 &   0.561 $\pm$  0.036 &  18.69 $\pm$  0.04 &  1.355 $\pm$ 0.068 &  0.498 $\pm$ 0.034 &  0.587 $\pm$ 0.029 &  0.326 $\pm$ 0.009 \\
  IPDF02 &  0.021 &   2.972 &   0.478 $\pm$  0.034 &  18.98 $\pm$  0.03 &  1.539 $\pm$ 0.077 &  0.612 $\pm$ 0.034 &  0.674 $\pm$ 0.034 &  0.393 $\pm$ 0.009 \\
  IPDF03 &  0.020 &   2.257 &   0.539 $\pm$  0.042 &  20.77 $\pm$  0.04 &  1.295 $\pm$ 0.065 &  0.566 $\pm$ 0.034 &  0.552 $\pm$ 0.028 &  0.360 $\pm$ 0.009 \\
  IPDF04 &  0.019 &   2.311 &   0.420 $\pm$  0.044 &  21.15 $\pm$  0.00 &  1.320 $\pm$ 0.066 &  0.616 $\pm$ 0.034 &  0.595 $\pm$ 0.030 &  0.390 $\pm$ 0.009 \\
\enddata
\tablecomments{$E(B-V)$ values are from \cite{SF11} obtained via the IPAC website. HI column densities are from the HI4PI survey \citep{HI4PI}. H$\alpha$ values are from the WHAM survey \citep{wham}.  The H$\alpha$ value for SCAL07 is contaminated by a nearby bright star. Dust temperatures are from \citep{dust_temp} and are derived using IRAS and Planck observations. FIR intensities and CIB measurements are based on Planck HFI observations \citep{planck2018}. FIR intensity values are the mean value within {\mbf the LORRI field of view} for the indicated target field. The CIB level has not been subtracted from the FIR intensity values given in columns 6 and 8. The CIB values given here are the mean value within the same {\mbf LORRI FOV} derived from the CIB maps created using the generalized needlet internal linear combination (GNILC) method \citep{needlet} for the indicated passband. We add to the zero-offset CIB GNILC maps the relevant monopole level from \citet{odegard}.}
\end{deluxetable}\label{tab:irpars}

\section{Decomposition of the Total Sky Intensities}\label{sec:decomp}

Given a total sky level for any field, our goal is to show whether or not we can account for all sources contributing to it. A portion of the sky should be due to the COB, and as part of the decomposition of the total sky, we will take as given the estimated IGL (integrated galaxy light) provided by all galaxies with the fields.  A second component is the light intensity foreground contributed by the integrated faint starlight (ISL) emitted by stars within the fields fainter than the photometric limit for detecting any single star directly.  We also have to account for scattered starlight light (SSL) falling in the the LORRI fields from bright stars {\it outside} the nominal field of view (bright {\it galaxies} outside the fields also contribute a minute amount of intensity). We also include the small contribution to the total sky from H$\alpha$ emission in the local interstellar medium.  Lastly, we must estimate the diffuse Galactic light (DGL) component contributed by light from the Milky Way scattered by dust in the interstellar medium into our line of sight. The COB is then defined to be the IGL estimate {\it plus} any anomalous intensity left over from this decomposition that cannot be attributed to a known source.

We do the decomposition in two steps.  The IGL, ISL, and SSL intensities are specified by external information, as discussed in detail in NH21 and NH22.  For convenience, we also provide brief summaries of these components in the subsections that follow. In passing, we also discuss the hydrogen ``two-photon" continuum correction that we included in NH22, but subsequently concluded was not needed.  Subtraction of the IGL, ISL, and SSL intensities from the total skies leaves the DGL and any anomalous component, which we then disentangle in a second step, as will be discussed in the following section.
\pagebreak
\subsection{Integrated Galaxy Light (IGL)}\label{sec:igl}

We compute the total IGL in two steps: the bright IGL {\mbf (hereafter BIGL)} for galaxies with $V < 19.9$ that were masked during the sky estimation process and the faint IGL for galaxies below this LORRI detection threshold. The IGL for the bright galaxies ($V < 19.9$) is estimated by extracting non-stellar objects in our LORRI field of view from the DESI/DECam Legacy Survey DR10 \citep{dls}. We use NOIRLab's Astro Data Lab interface for retrieving the galaxy catalogs. The transformation to $V$-mag from the DLS $g, r$ bands (or the $g, i$ bands when $r$-band is unavailable) is derived from eight templates of galaxy spectral energy distributions spanning the morphologies E, S0, Sa, Sb, Sc, and Ir types. We weight the templates by the morphological fractions observed in the field population of galaxies and derive an average $(V-g)$ vs. $(g-r)$ relationship (or vs. $(g-i)$ when $r$-band info is unavailable) over the redshift range $0 < z \le 1$, typical for brighter galaxies. Our best-fit color transformations are
\begin{equation} \label{eqn:Vtrans}
\begin{split}
(V-g) & = 0.6216 - 2.03(g-r) + 0.7221(g-r)^2 \\
(V-g) & = 2.265 - 4.557(g-i) + 2.238(g-i)^2 - 0.3624(g-i)^3
\end{split}
\end{equation}
These relations are valid over the range $0.6 \leq (g-r) \leq 1.5$ and $0.4 \leq (g-i) \leq 2.8$.
We then derive the IGL intensity contribution based on the $V$ magnitude and sum up the contributions for all DLS galaxies with $V < 19.9$ in the LORRI field of view. The statistical error for the bright IGL is $\sim 0.01 {\rm~nW~m^{-2}~sr^{-1}}$ and is derived from the photometric errors given in the DLS catalogs. The systematic error for the bright IGL, typically $\sim 0.07 {\rm~nW~m^{-2}~sr^{-1}}$, is derived from the uncertainties associated with the color transformation to the $V$-band. 
{\mbf As noted in Table~\ref{tab:program}, the BIGL addition for galaxies brighter than the LORRI detection limit was not applied in our NH21 analysis but was applied in NH22 and is applied in this work. The typical BIGL component for the fields used in NH21 is $\rm \sim1.2~nW~m^{-2}~sr^{-1}$.}

The precepts for estimating the faint IGL due to galaxies at or below the $V=19.9$ detection threshold are discussed at length in NH21. We estimate the uncertainty in the faint IGL intensity by assessing the specific contribution to the error from the systematic terms (errors in the fits to the galaxy number counts) and from the statistical errors (cosmic variance). The two systematic errors associated with the fits to the galaxy number counts are from the errors in the coefficients to the power law fits used in NH21 and the error associated with the form of the fitting function. The formal errors in the power law coefficients yield a fractional error of 13.1\% in the IGL intensity. The difference between the IGL derived from the power law fits versus that derived using a quadratic fit to the galaxy counts yields a fractional change in the IGL of 6.6\%. Summing these two error components in quadrature yields a combined systematic fractional error of 14.7\% in the IGL intensity.
The total error in the IGL must also include the statistical uncertainty due to the effects of cosmic variance over a single LORRI field-of-view (FOV). The cosmic variance error for a single LORRI FOV used in this work is the same as the single-field CV error adopted in NH21 \citep{cosvar2008} - which translates to an IGL fractional error of 11.8\%. Summing, in quadrature, this statistical error with the above systematic error yields a total fractional error of 18.8\% in the faint IGL intensity. 
Combining the bright and faint galaxy contributions to the IGL gives a total IGL intensity for each of our survey fields. This IGL corresponds to the expected light in the LORRI bandpass from all galaxies brighter than $V=30$ mag. The bright and faint IGL values and their associated total errors are provided in Table~\ref{tab:skyfluxes} for each field.

\subsection{Scattered Light from Bright Stars (SSL) and Galaxies (SGL)}

The LORRI instrument accepts scattered starlight from sources well outside its field of view. The NCOB fields were selected to minimize SSL, and the SSL intensity remaining for any given field depends on the specific stars that surround it. As detailed in NH21, the LORRI scattered light function is estimated from pre-launch calibration tests, and in-flight measurements of the scattered {\it sunlight} background as a function of angular distance from the Sun.  The SSL is estimated from the convolution of the scattering function with stars surrounding the field as provided by the Tycho2 star catalog \citep{tycho2}, and the Yale Bright Star catalog v5.0 \citep{YBSC5} for bright stars, with fainter stars ($11 \le V < 20$ mag) provided by the {\it Gaia} DR3 catalog \citep{gaia2016,gaia2023}. The error in the SSL intensity {\tbf reflects the 10\% scatter in} the LORRI scattering function, and thus {is systematic} over all fields. We used the Gaia ESA Archive to retrieve all of the above star catalogs for each field.  The SSL values for each field are given in Table~\ref{tab:skyfluxes}.

The SGL term is the analogous scattered light contributed by bright {\it galaxies} outside the LORRI field. {\mbf As with the SSL calculation, the contribution to the SGL is calculated out to an off-axis angle of $45^\circ$. Because no uniform all-sky galaxy catalog yet exists to perform this calculation using the actual positions and fluxes of known galaxies, we estimate the SGL as described in NH21. Briefly, we use the galaxy number counts from well-calibrated surveys to compute the mean surface brightness of galaxies with $V<20$ and then compute the contribution to each annular bin extending out to a radius of $45^\circ$. The flux contributions in each bin are convolved with the LORRI scattering function and are then summed up to provide the final SGL estimate. The same SGL value is adopted for all fields. The surface density of bright galaxies is so low that this intensity, $0.10\pm0.01{\rm  ~nW ~m^{-2} ~sr^{-1}},$ is almost negligible. Hence, even using actual galaxy positions and brightnesses would not make a substantial difference in the final results. As with the SSL, the uncertainty in SGL is taken to be 10\%.}

As noted in $\S\ref{sec:design},$ we observed eight ``SCAL" calibration fields to test the SSL corrections. Figure \ref{fig:scal} shows the predicted SSL corrections for the NCOB and SCAL fields as compared to the {\it inferred} SSL corrections estimated by subtracting all other intensity components from the total observed sky intensity.  In detail, this means subtracting the field-specific IGL, SGL, ISL, and H$\alpha$ components discussed in this section, as well as the DGL corrections (which will be discussed in the following section), {\mbf such that SSL$_{inf}$ = TotSky - IGL - ISL - H$\alpha$ - DGL - $\rm S_U$.}
Note that we also subtracted the small anomalous $\rm S_U$ intensity component.  This is constant over all fields, and will be discussed in detail in $\S\ref{sec:cob}.$ 

As is seen in the figure, there is excellent agreement between the predicted and inferred SSL intensities for the 16 NCOB fields.
{\tbf The three SCAL fields with the largest predicted SSL values, however, have relatively much smaller inferred SSL values.  These fields were selected to return strong SSL backgrounds by positioning the LORRI field-of-view close to bright stars.  The implication is that the scattered light function $\lesssim0\adeg5$ radii from the LORRI field over estimates the scattered light contribution. In contrast, the NCOB fields were positioned to avoid angularly-close bright stars; the fainter stars remaining close to those fields contribute little to the total SSL integral. Likewise, the five SCAL fields with the smaller predicted SSL intensities also agree well with their corresponding inferred SSL backgrounds.} 

Based on the SCAL data, we do not see a strong case for changing the present SSL estimation procedure, which appears to work well for the SSL intensity range experienced by our NCOB and DCAL fields.  {\tbf Comparing the predicted and inferred SSL values directly for the 16 NCOB fields, we find that the rms relative difference between the two values to be 15\%.  As the error in the inferred SSL values strongly dominates in this comparison, it means that the error in the {\it predicted} SSL value must be yet smaller. The five SCAL fields close to the equality line in Figure \ref{fig:scal} in aggregate have markedly higher SSL backggrounds than do the NCOB fields, but here the rms relative difference between the inferred and predicted SSL values rises only slightly to 18\%.}

\begin{figure}[hbtp]
\centering
\includegraphics[keepaspectratio,width=6.0 in]{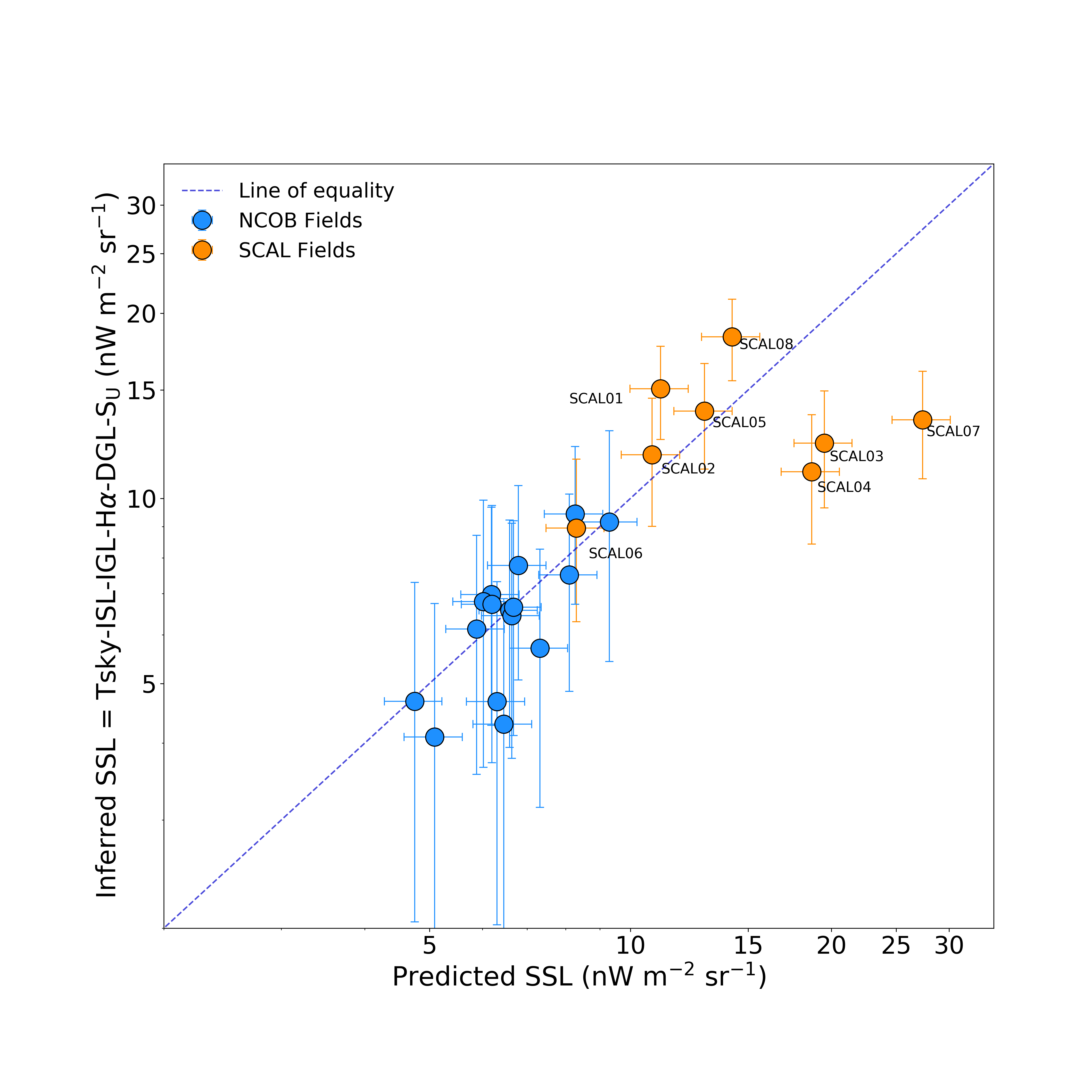}
\caption{The inferred SSL as a function of the predicted SSL for the 16 COB fields (blue points) and the 8 SCAL fields (orange points). The predicted SSL is the intensity derived using the LORRI large-angle PSF convolved with  known stars from various catalogs. The inferred SSL is the residual signal when all other non-SSL intensity components are subtracted from the total observed sky intensity. {\tbf A log-log scale is used as the SSL error is proportional to the SSL signal.} The dashed line shows the line of equality. The three outlier SCAL fields all have more than one $V < 9.2$ star (intentionally) located within 0.32 degrees of the LORRI field center.}
\label{fig:scal}
\end{figure}

\subsection{Integrated Faint Starlight (ISL)}

ISL is the integrated light of faint Galactic stars {\it within} any field that are fainter than the LORRI photometric detection limit. Our approach is to integrate {\tt TRILEGAL} models \citep{tril2005, trilv16} of the expected population of faint stars down to $\rm V = 30$ within a $1~\rm deg^2$ region centered on our fields, following the procedures presented in NH21. For the present fields, the bright limit of the intensity integral (Eq. 3 in NH21) is $V=19.9.$  The ISL errors (Table \ref{tab:skyfluxes}) are a combination of systematic and random errors due to uncertainties in the {\tt TRILEGAL} model parameters and the estimated fluctuations in the star counts, respectively. As seen in Figure~\ref{fig:gaia_vs_trilegal}, the {\tt TRILEGAL} predictions agree very well with the actual star counts from {\it Gaia} DR3 \citep{gaia2016, gaia2023}. We thus have confidence that the use of the {\tt TRILEGAL} for predicting the integrated faint starlight contribution down to $\rm V=30$ is a reliable approach. The ISL values for each field are given in Table~\ref{tab:skyfluxes}.

\begin{figure}[hbtp]
\centering
\includegraphics[keepaspectratio,width=7.0 in]{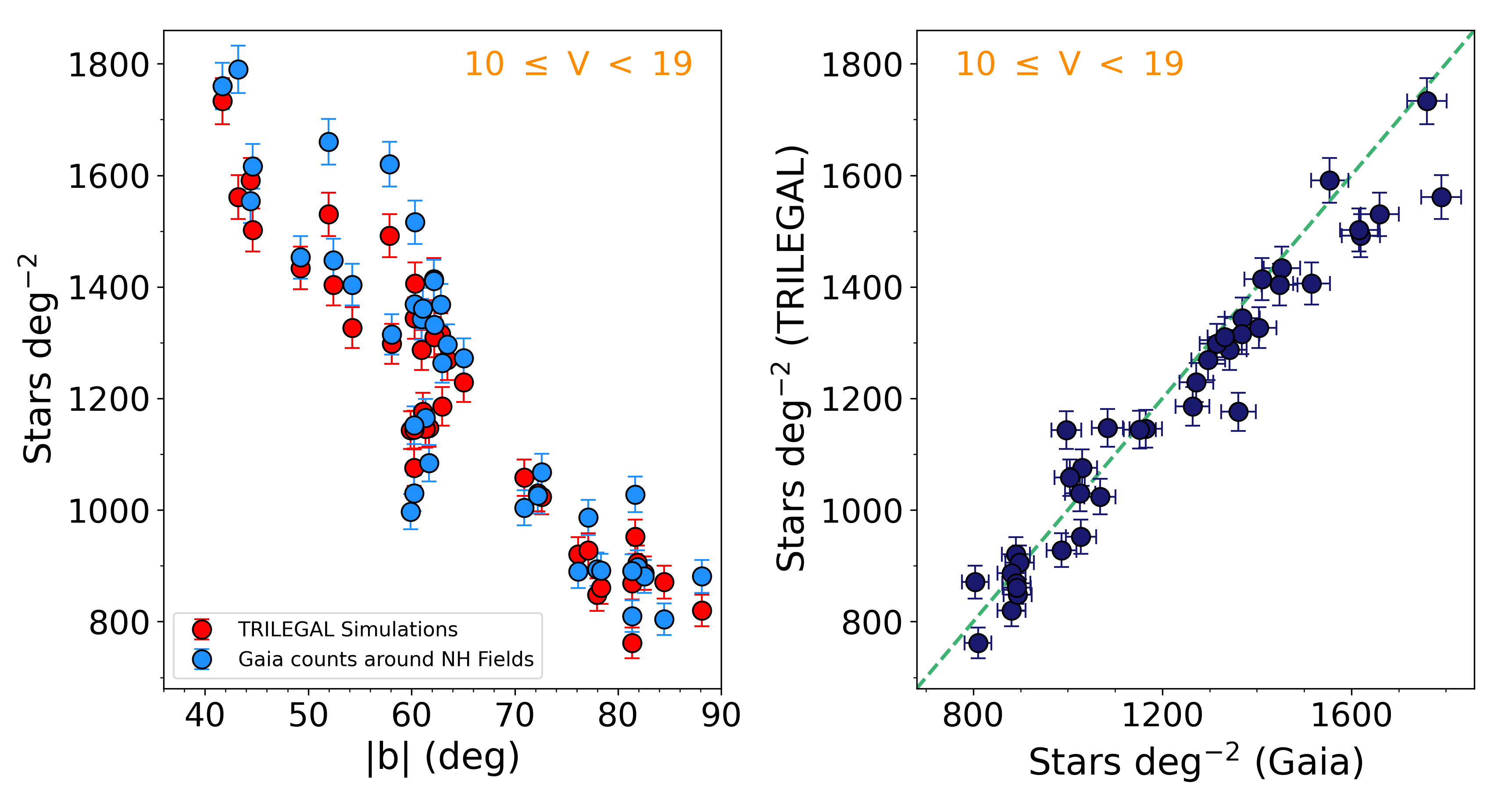}
\caption{The left-hand plot shows the stellar density on the sky as a function of the absolute value of Galactic latitude from {\it Gaia} DR3 and from the {\tt TRILEGAL} Milky Way model simulator for each of the 32 fields in the current work plus the 7 fields from NH21. To enable the comparison with {\it Gaia}, the stellar density calculations are limited to stars in the range $\rm 10 \leq V < 19$. The right-hand plot shows the direct comparison of star counts from {\it Gaia} vs. {\tt TRILEGAL} for this same magnitude range. The dashed green line represents the line of equality.}
\label{fig:gaia_vs_trilegal}
\end{figure}

\subsection{The Two-Photon Continuum and H$\alpha$ Foregrounds}

The existence of a full-sky diffuse Ly-$\alpha$ background from the Milky Way \citep[e.g.][]{gladstone} {\tbf suggests that there might be } associated hydrogen two-photon continuum \citep[][2PC]{two} at some sky locations.  In NH22, we noted that the preliminary analysis of deep {\tbf 130-180 nm} UV-spectra taken of NHTF01 suggested {\tbf that a significant fraction of the continuum in the UV was due to 2PC. The corresponding optical intensity in the LORRI passband was estimated to be $0.93\pm0.47 {\rm~nW~m^{-2}~sr^{-1}}$.} {\tbf Subsequent analysis of the NHTF01 UV-spectra greatly reduced the likelihood that 2PC was needed to account for the observed continuum.  The strongest argument against this hypothesis, however, was provided by the extremely low level of H$\alpha$ emission seen over the LORRI field. As such, in contrast to the analysis in NH22, we conclude that 2PC does not contribute to the observed sky levels at anything more than a trivial level.}

{\tbf The generation of 2PC emission occurs during the recombination of ionized hydrogen.  The associated H$\alpha$ emission is a direct predictor of 2PC continuum intensity \citep{kuv, ksuv}. 
The H$\alpha$ levels seen in the COB survey fields (see Table \ref{tab:irpars}) are {\tbf all} very low ($<1$ R) based on observations made with the Wisconsin H$\alpha$ Mapper (\cite{wham}; WHAM). This in turn implies that any 2PC emission present should have intensity $< 0.1 {\rm~nW~m^{-2}~sr^{-1}}$.} We conclude that 2PC emission is negligible in the present NCOB fields.}

{\tbf That said, since H$\alpha$ emission is present in its own right in the LORRI fields, it will contribute directly to the observed total sky level, albeit at an exceedingly modest level.} For one Rayleigh of H$\alpha$ emission, the associated intensity is ${\rm 0.24 ~nW ~m^{-2} ~sr^{-1}};$ the median H$\alpha$ intensity for the NCOB and DCOB fields is about 2/3 of that (see Table~\ref{tab:skyfluxes}). 

\begin{figure}[hbtp]
\centering
\includegraphics[keepaspectratio,width=7.0 in]{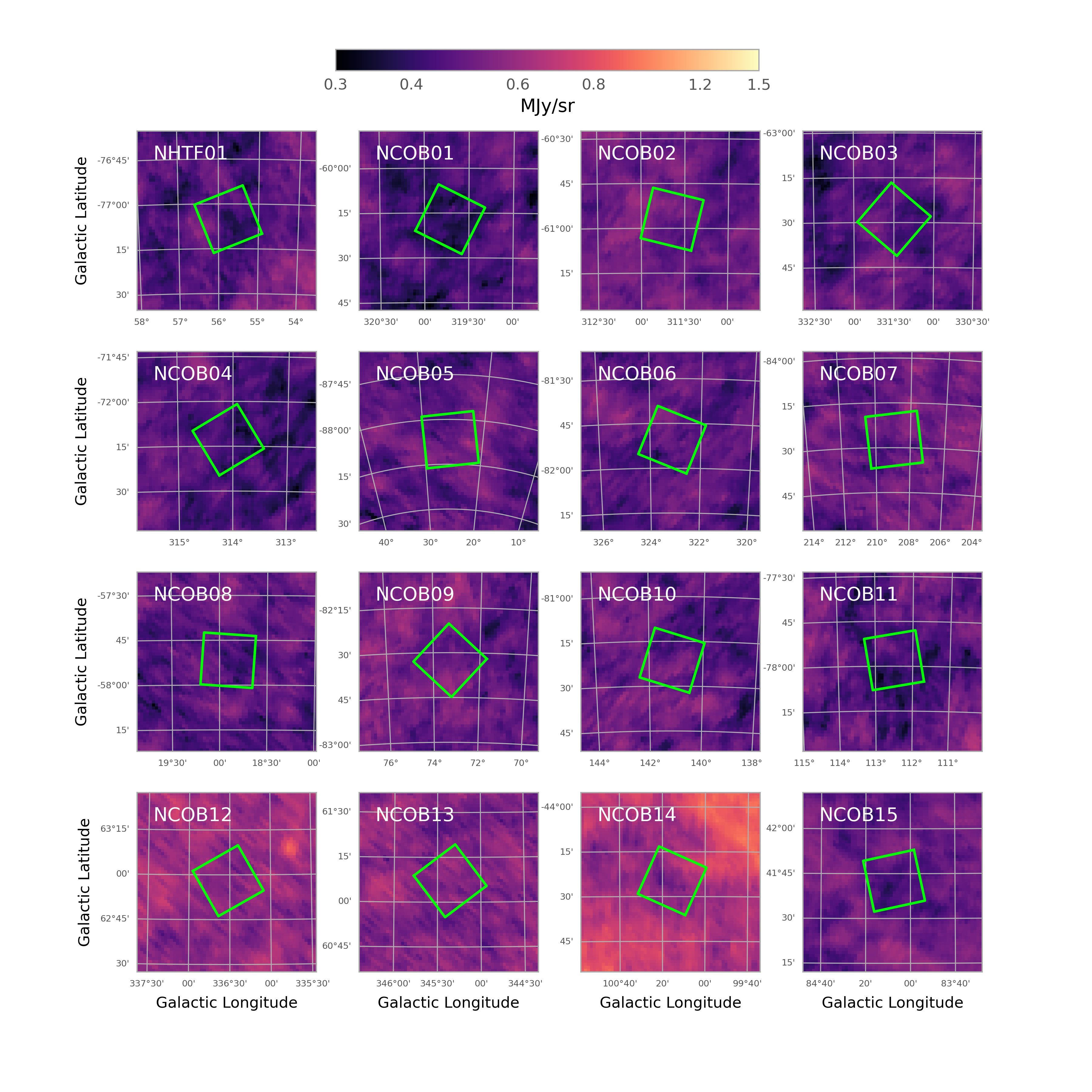}
\caption{The locations of the NCOB fields are shown with respect to the 545 GHz ($550~{\rm\mu m}$) Planck HFI map \citep{planck2018}. The field of view of each HFI cutout image is $1^{\circ}~\times~1^{\circ}$. The green boxes show the LORRI CCD orientation relative to Galactic coordinates. }
\label{fig:550_fields}
\end{figure}

\section{The Diffuse Galactic Light Foreground}\label{sec:dgl}

The approach to estimating the DGL contribution to the sky level in any field is to first estimate the amount of interstellar dust along the line of sight, and then calculate the intensity of Milky Way light that it would scatter. In NH21 and NH22, we used the DGL estimator of \citet{zemcov}, which uses the residual $100{~\rm\mu m}$ intensity above the cosmic infrared background (CIB) to estimate the amount of IR cirrus in the field, and scattering theory to estimate the actual DGL scattered into the line of sight.  Unfortunately, uncertainties in the needed theoretical parameters translate into large ($\sim40\%$) errors in the DGL estimates. \citet{zemcov} suggested that better accuracy might be obtained by observing fields over a range of $100{~\rm\mu m}$ intensity in order to derive an empirical relation between the DGL level and the FIR (far-infrared) indicator intensity. This approach was later attempted by \citet{symons} in their independent measurement of the COB intensity from NH archival LORRI observations.

As discussed in $\S\ref{sec:design},$ we used $100{~\rm\mu m}$ intensity to select both the NCOB science fields and the DGL calibration fields, and had planned to derive an empirical DGL estimator based on the average $100{~\rm\mu m}$ intensity in any given field. In the initial phases of this work, however, we discovered an error in our previous DGL estimates. As detailed in NH21, we concluded that the IRIS $100{~\rm\mu m}$ map included an amount of residual zodiacal light, and thus derived a correction to the input $100{~\rm\mu m}$ intensity as a function of ecliptic latitude of the fields (see Figure 16 of NH21). We now understand that the rise in intensity with decreasing (absolute) ecliptic latitude is due an extended zone of high Galactic latitude dust emission that overlaps the ecliptic plane, which we simply missed in our earlier analysis. We thus no longer apply a residual zodiacal light correction to our FIR input intensities, although we do still retain a $|\beta|<15^\circ$ exclusion zone for COB fields, given concern with potential systematic effects in the zodiacal light corrections.  We also note that this error caused us to under-estimate the amplitude of the DGL foreground in NHTF01 analysed in NH22; we rework the decomposition of that field as part of the present analysis.

In the course of re-developing a $100{~\rm\mu m}$-based DGL estimator, we also explored the utility of other FIR bands to predict the mass density of IR cirrus in our fields.  We found that the $350{~\rm\mu m}$ (857 GHz) and $550{~\rm\mu m}$ (545 GHz) intensities from the Planck High Frequency Instrument (HFI; \cite{planck2014, planck2016}) both predict the DGL foreground with significantly higher precision. Further improvements were obtained by using $350{~\rm\mu m}$ and $550{~\rm\mu m}$ intensities in combination.  The result is a strong improvement in characterizing the amplitude of DGL in the total sky signals for any given field. We will demonstrate this at the end of the section by comparing single-band DGL-estimators at 100, 350, 550, and $849{~\rm\mu m}$ to the final $350{~\rm\mu m}$ plus $550{~\rm\mu m}$ relation. Figure~\ref{fig:550_fields} shows the overlay of the LORRI field of view for our 16 COB science fields on the $550~\mu$m HFI images. While our our NCOB fields were initially selected to minimize DGL based on the $100~\mu$m emission, the $550~\mu$m emission in each NCOB field is also very low, further confirming the selection of these fields as optimized pointings for measuring the COB.

\subsection{The Use of FIR Background Intensities to Estimate DGL}

Both the NCOB and DCAL fields were selected to have relatively low IR cirrus optical depth. Thus for dust at a given temperature we expect its emitted thermal FIR intensity to be linearly related to its surface density. The corresponding optical DGL intensity is thus proportional to the dust surface density, but it is moderated by a geometric factor, $g(b),$ that accounts for changes in the scattering phase angle with Galactic latitude. 
In terms of the total sky, $S_T,$ and the intensity components discussed in the previous section, we define the DGL+:
\begin{equation}\label{eqn:IDGL}
{\rm DGL+ = S_T-IGL-SSL-SGL-ISL-I(H\alpha)},
\end{equation}
where DGL+ is the optical intensity that remains in a given field after all other known signals are subtracted from the field's total sky. The terms on the right hand side of Equation~\ref{eqn:IDGL} are all based on the measurements made in the optical passband. We refer to this as DGL+ because it consists of the DGL signal plus any anomalous intensity from sources presently not known or considered.
For a single FIR band, 
a DGL+ estimator can be made by performing a least squares linear fit using as the independent variable the FIR intensity, $I(\lambda),$ corrected for the cosmic background intensity at the wavelength of the FIR intensity, $CIB(\lambda)$ {\tbf \citep{symons}}. In detail:
\begin{equation}\label{eqn:dgl_pred}
{\rm DGL+} = a~g(b)~\bigg(I(\lambda) - CIB(\lambda)\bigg) +c,
\end{equation}
where $g(b)$ is the geometric scaling factor \citep{jura, zemcov} which is defined 
as
\begin{equation}
g(b)=\frac{1-0.67\sqrt{\sin|b|}}{0.376}, \label{eqn:geo}
\end{equation}
where $b$ is the Galactic latitude. In this definition $g(b)$ is normalized to be  unity at $|b|= 60^\circ,$  and thus only describes the relative effects of latitude.
The absolute conversion from FIR intensity to optical DGL is provided by the coefficient, $a,$ in equation~(\ref{eqn:dgl_pred}) which is determined empirically by fitting the independent variable $(g(b)\times(I(\lambda) - CIB(\lambda)))$ to the dependent variable $(S_T-IGL-SSL-SGL-ISL-I(H\alpha))$ over the combined sample of NCOB and DCAL fields. The intercept, $c,$ provides an estimate of any anomalous intensity contribution.   

\subsection{FIR Intensities and the Cosmic Infrared Background}\label{sec:FIRCIB}

We used the Planck 2018 HFI maps \citep{planck2018} available from the NASA/IPAC Infrared Science Archive (IRSA) to provide the 350, 550, and $849{~\rm\mu m}$ FIR intensities.  As is evident in Figure \ref{fig:550_fields}, the LORRI field is large enough for the FIR background to vary over its extent.  {\mbf We thus took care to compute the average intensity within the full LORRI field of view.} Those average FIR intensities for each field are tabulated in Table \ref{tab:irpars}. The errors in the FIR intensities are predominantly systematic as the $\sim5$\% calibration uncertainties in HFI are the dominant source of uncertainty \citep{planck2016}.

For the low FIR emission fields selected, the CIB and dust thermal intensities are in rough parity. As such, accurate knowledge of the CIB is critical to correctly isolating the dust thermal emission needed to estimate the optical DGL foreground.  The CIB has two components.  The first is effectively the ``intensity monopole" or full-sky average level of the background over the sky after removing dust and other Milky way emission sources.  The second is the field-to-field variations in the background due to CIB anisotropies.

For the monopole intensities at 350, 550, and $849{~\rm\mu m},$ we draw on \citet{odegard}, who derive  $0.576\pm0.034,$ $0.371\pm0.018,$ and $0.149\pm0.017~{\rm MJy ~sr^{-1},}$ respectively, from Planck HFI maps \citep{planck2014, planck2016}. Briefly, they use H~I column density as an indicator of dust surface density, which for low dust optical depth is linearly correlated with FIR intensity.  The CIB monopole is derived by extrapolating this relation to zero H~I column density.  As the CIB monopole intensities are common to all fields, the errors in the intensities are treated as systematic. 

\begin{figure}[hbtp]
\centering
\includegraphics[keepaspectratio,width=7.0 in]{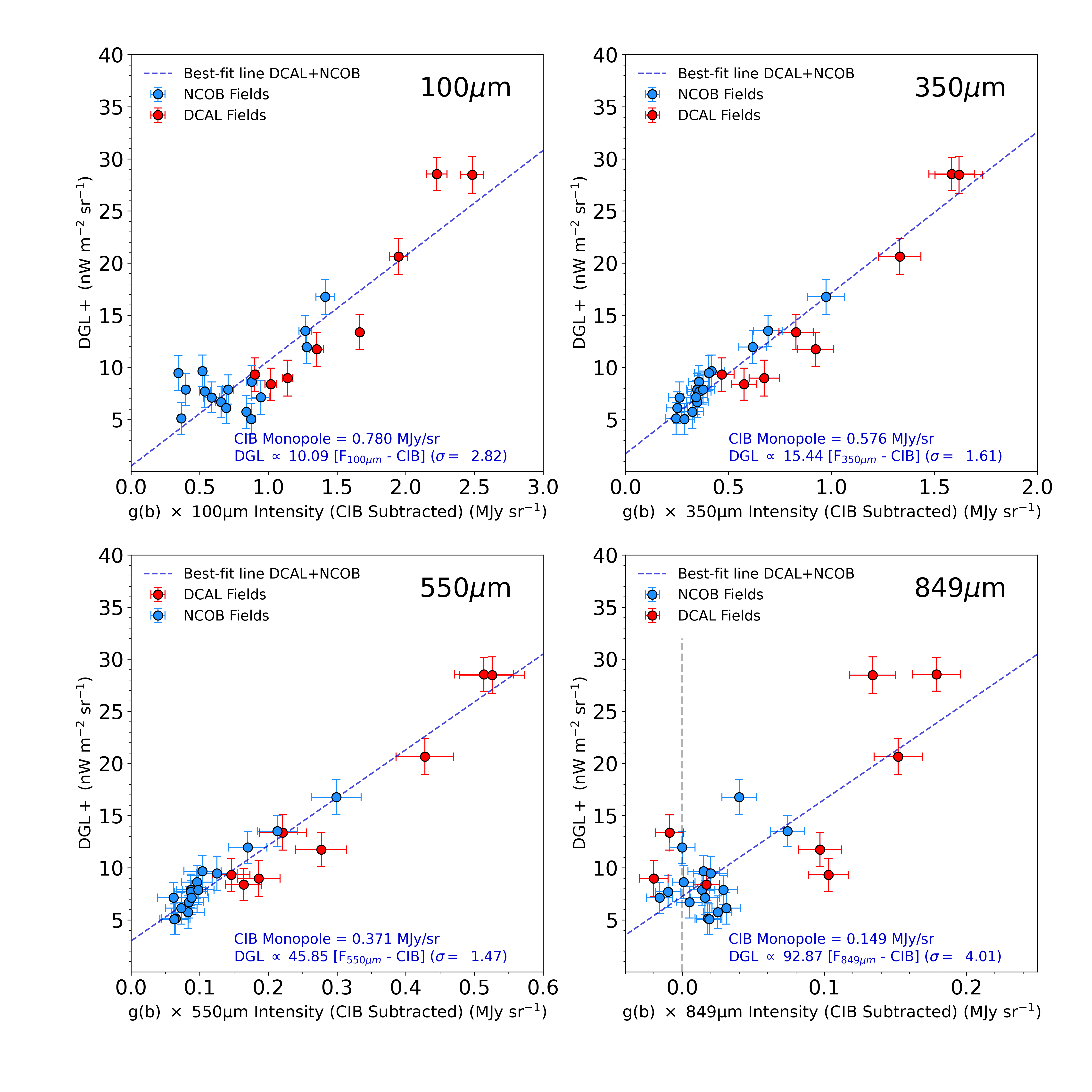}
\caption{The relationships between DGL+ intensity in the NCOB and DCAL fields as a function of IRIS $100{~\rm\mu m},$ Planck HFI $350{~\rm\mu m},$ $550{~\rm\mu m},$ or $849{~\rm\mu m}$ intensity averaged over the LORRI field are shown. The CIB background intensity has been subtracted from the input FIR intensities to isolate the FIR emission from dust alone. The $g(b)$ is a function that accounts for the phase-angle dependence of the dust scattering cross section on Galactic latitude (see eqn. \ref{eqn:geo}).}
\label{fig:dgl_single}
\end{figure}

The CIB anisotropies for any field are provided by the \citet{needlet} analysis of the Planck HFI maps, which uses a form of power-spectrum analysis, referred to as the generalized needlet internal linear combination (GNILC) method, to separate the structure of the anisotropies from that of the thermal dust emission. These are treated as a field-dependent intensity correction to the overall CIB monopole. Specifically, the CIB-subtracted FIR intensity for a given field is 
\begin{equation}\label{eqn:cibsubflux}
    I_c(\lambda) = I(\lambda) - CIB_{GNILC}(\lambda) - \rm CIB\_MONOPOLE,
\end{equation}
where $I(\lambda)$ and $CIB_{GNILC}(\lambda)$ are the averages over {\mbf the LORRI field} from the HFI and GNLIC CIB maps hosted at IRSA and CIB\_MONOPOLE is the relevant \citet{odegard} value.  The anisotropy corrections, $CIB_{GNILC}(\lambda) + \rm CIB\_MONOPOLE,$ are tabulated in Table \ref{tab:irpars}; the field-to-field variations of the CIB values are treated as random errors. In application, we have found that accounting for the CIB intensity anisotropies in this way removes a significant source of variance in the relations between FIR and DGL+.

Lastly, the Planck HFI did not observe at $100{~\rm\mu m}.$
We thus rely on the IRIS reprocessing of the IRAS full-sky thermal-IR maps \citep{iris} for intensities at this wavelength. The CIB intensity at this wavelength is taken to be {\tbf$0.78\pm0.21~{\rm MJy ~sr^{-1}}$ \citep{lagache}.}

\subsection{The DGL Estimators}

Figure \ref{fig:dgl_single} shows the four single-band DGL estimators based on 100, 350, 550, and $849{~\rm\mu m}$ intensities, derived by fitting equation (\ref{eqn:dgl_pred}) to the NCOB and DCAL DGL+ values (equation~\ref{eqn:IDGL}). {\mbf The linear fit parameters are given in Table~\ref{tab:dgl_fir_fits}.} The slope of the $100{~\rm\mu m}$ estimator is nearly the same as that in the \citet{zemcov} theoretical estimator, but clearly has much smaller errors.  The $350{~\rm\mu m}$ and $550{~\rm\mu m}$ estimators have even tighter trends, with the scatter in the $550{~\rm\mu m}$ estimator nearly a factor of two smaller than that in the $100{~\rm\mu m}$ estimator.  In contrast, the $849{~\rm\mu m}$ estimator offers the poorest performance of the four bands tested.

\begin{deluxetable}{ccccccc}
\tabletypesize{\small}
\tablecolumns{7}
\tablewidth{0pt}
\tablecaption{\mbf Linear Fit Parameters to Single-band FIR Intensity - DGL+ Relations}
\tablehead{
\multicolumn{7}{c}{~}\\
\multicolumn{1}{c}{Wavelength} &
\multicolumn{1}{c}{Slope} &
\multicolumn{1}{c}{Intercept} &
\multicolumn{1}{c}{Fit RMS} &
\multicolumn{1}{c}{R} &
\multicolumn{1}{c}{CIB Monopole} &
\multicolumn{1}{c}{CIB} \\
\multicolumn{1}{c}{($\mu$m)} &
\multicolumn{1}{c}{$\rm (nW~m^{-2}~sr^{-1})/(MJy~sr^{-1})$} &
\multicolumn{1}{c}{$\rm (nW~m^{-2}~sr^{-1})$} &
\multicolumn{1}{c}{$\rm (nW~m^{-2}~sr^{-1})$} &
\multicolumn{1}{c}{Value} &
\multicolumn{1}{c}{$\rm (MJy~sr^{-1})$} &
\multicolumn{1}{c}{Reference}
}
\startdata
  100 & 10.09 $\pm$ 0.59 & 0.53 $\pm$ 0.68 & 2.82 & 0.895 & 0.780 $\pm$ 0.210 & \cite{lagache} \\
  350 & 15.44 $\pm$ 0.84 & 1.71 $\pm$ 0.58 & 1.61 & 0.968 & 0.576 $\pm$ 0.034 & \cite{odegard} \\
  550 & 45.85 $\pm$ 2.48 & 2.97 $\pm$ 0.53 & 1.47 & 0.973 & 0.371 $\pm$ 0.009 & \cite{odegard} \\
  849 & 92.87 $\pm$ 6.22 & 7.26 $\pm$ 0.39 & 4.01 & 0.785 & 0.149 $\pm$ 0.006 & \cite{odegard} \\
\enddata
\tablecomments{\mbf The intercept values are dependent on the CIB monopole values, which are thus listed here (see $\S\ref{sec:FIRCIB}$). The slope of the FIR intensity - DGL+ relationship is independent of the CIB monopole value. These linear fits are provided for reference only. We do not use them in deriving the estimate of the COB in this work.}
\end{deluxetable}\label{tab:dgl_fir_fits}

One possible reason for larger scatter in the estimators at $100{~\rm\mu m}$ vs. $550{~\rm\mu m}$ is field-to-field variation in the dust temperature. The average dust temperature for the NCOB and DCAL fields is $20.1$ K, with a dispersion of $1.9$ K.  Variation in dust temperature can cause variations in the strength of the FIR intensity emitted for the same surface density of dust and the same optical light scattered. At $20$ K, $100{~\rm\mu m}$ falls on the short wavelength side of the peak of the black-body spectrum, and thus $100~\mu$m emission is more sensitive to small temperature changes than is the intensity at $550{~\rm\mu m},$ which falls on the long wavelength side of the peak. 
We thus investigated two-band DGL estimators, finding that using the $350{~\rm\mu m}$ and $550{~\rm\mu m}$ intensities in combination gives the best performance, returning smaller scatter than the $550{~\rm\mu m}$ single-band estimator.

\begin{figure}[hbtp]
\centering
\includegraphics[keepaspectratio,width=6.0 in]{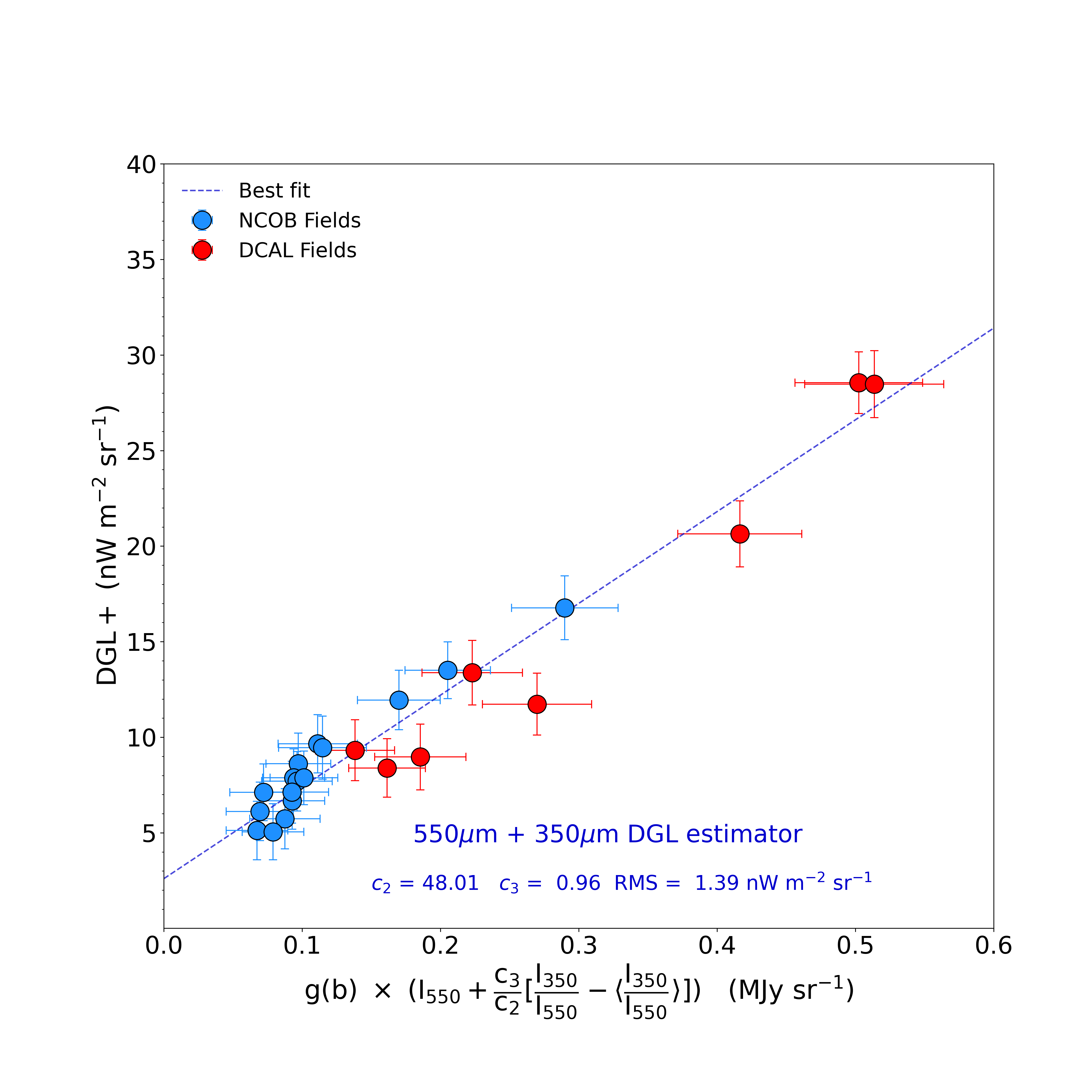}
\caption{This DGL estimator uses both Plank HFI $350{~\rm\mu m}$ and $550{~\rm\mu m}$ background intensities as input.  The $550{~\rm\mu m}$ intensity is modified by the residual of the ratio of $350{~\rm\mu m}$ to $550{~\rm\mu m}$ intensity about the mean value of their ratio.}
\label{fig:dgl_3par}
\end{figure}

In detail, the two-band estimator fits the DGL+ value for any field as:
\begin{equation}\label{eqn:2-band}
\rm DGL+ = c_1 + g(b)~\bigg[c_2 I_c(550{~\rm\mu m})+c_3 \bigg(\frac{I_c(350{~\rm\mu m})}{I_c(550{~\rm\mu m})}~-~\left\langle \frac{I_c(350{~\rm\mu m})}{I_c(550{~\rm\mu m})} \right\rangle\bigg)\bigg],
\end{equation}
where the subscripts on the intensities indicate that the CIB intensity has been subtracted from them. The mean $350{~\rm\mu m}$ to $550{~\rm\mu m}$ intensity ratio is subtracted from ratio for each field to strongly reduce covariance of the intensity-ratio term with the overall intercept term. The fit of this estimator is shown in Figure \ref{fig:dgl_3par}, with coefficient values {\mbf  $c_1=2.60,$ $c_2=48.01,$ and $c_3=0.96,$ with rms residuals of ${\rm 1.39~nW ~m^{-2} ~sr^{-1}}.$ The mean $350{~\rm\mu m}$ to $550{~\rm\mu m}$ intensity ratio for our sample is 3.66.}

To use the two-band estimator to predict just the DGL value, we subtract the $c_1$ coefficient, as the predicted DGL must go to zero when the FIR intensity goes to zero. Specifically, our DGL predictor is:
\begin{equation}\label{eqn:dgl_predictor}
\mbf \rm DGL({\rm nW~m^{-2}~sr^{-1}}) = g(b)~\bigg[48.01~I_c(550{~\rm\mu m})~+~
0.96~\bigg(\frac{I_c(350{~\rm\mu m})}{I_c(550{~\rm\mu m})}~-~3.66\bigg)\bigg],
\end{equation}
where the FIR intensities are in units of MJy sr$^{-1}$ and the predicted DGL value is in units of ${\rm nW~m^{-2}~sr^{-1}}$. The estimated DGL values for all {\mbf NCOB and DCAL} fields are given in Table~\ref{tab:skyfluxes}. {\mbf The errors in Table~\ref{tab:skyfluxes} are the quadrature sum of the random and systematic errors. Those individual random and systematic errors are given in Table~\ref{tab:errorbudget}. The errors for DGL are the rms values obtained from the Monte Carlo analysis described in $\S\ref{sec:cob}$}.

\begin{deluxetable}{lcccccrcc}
\tabletypesize{\scriptsize}
\tablecolumns{9}
\tablewidth{0pt}
\tablecaption{Sky Component Intensity Levels for Each NH COB and DCAL Field}
\tablehead{
\multicolumn{9}{c}{~} \\
\multicolumn{2}{c}{~} &
\multicolumn{2}{c}{Scattered Light from} &
\multicolumn{1}{c}{Faint} &
\multicolumn{1}{c}{H$\alpha$} &
\multicolumn{1}{c}{Diffuse} &
\multicolumn{1}{c}{Faint} &
\multicolumn{1}{c}{Bright}\\
\multicolumn{1}{c}{Field} &
\multicolumn{1}{c}{Total Sky} &
\multicolumn{1}{c}{Stars} &
\multicolumn{1}{c}{Galaxies} &
\multicolumn{1}{c}{Stars} &
\multicolumn{1}{c}{Intensity} &
\multicolumn{1}{c}{Galactic} &
\multicolumn{1}{c}{Galaxies} &
\multicolumn{1}{c}{Galaxies} \\
\multicolumn{1}{c}{ID} &
\multicolumn{1}{c}{($\rm S_T$)} &
\multicolumn{1}{c}{(SSL)} &
\multicolumn{1}{c}{(SGL)} &
\multicolumn{1}{c}{(ISL)} &
\multicolumn{1}{c}{(I(H$\alpha$))} &
\multicolumn{1}{c}{Light (DGL)} &
\multicolumn{1}{c}{(IGL)} &
\multicolumn{1}{c}{(BIGL)}
}
\startdata
  NHTF01 &  22.82 $\pm$  0.77 &   6.19 $\pm$  0.62 &   0.10 $\pm$  0.01 &   1.16 $\pm$  0.18 &   0.12 $\pm$  0.01 &   4.54 $\pm$  0.95 &   6.61 $\pm$  1.24 &   1.83 $\pm$  0.12 \\
  NCOB01 &  23.15 $\pm$  0.50 &   6.59 $\pm$  0.66 &   0.10 $\pm$  0.01 &   1.78 $\pm$  0.22 &   0.17 $\pm$  0.01 &   4.46 $\pm$  1.04 &   6.61 $\pm$  1.24 &   1.17 $\pm$  0.07 \\
  NCOB02 &  24.99 $\pm$  0.54 &   6.79 $\pm$  0.68 &   0.10 $\pm$  0.01 &   1.63 $\pm$  0.21 &   0.19 $\pm$  0.01 &   5.28 $\pm$  1.14 &   6.61 $\pm$  1.24 &   1.63 $\pm$  0.16 \\
  NCOB03 &  22.93 $\pm$  0.63 &   6.64 $\pm$  0.66 &   0.10 $\pm$  0.01 &   1.66 $\pm$  0.21 &   0.21 $\pm$  0.01 &   4.60 $\pm$  1.01 &   6.61 $\pm$  1.24 &   1.47 $\pm$  0.06 \\
  NCOB04 &  20.68 $\pm$  0.50 &   7.32 $\pm$  0.73 &   0.10 $\pm$  0.01 &   1.27 $\pm$  0.18 &   0.24 $\pm$  0.01 &   3.30 $\pm$  0.93 &   6.61 $\pm$  1.24 &   1.16 $\pm$  0.02 \\
  NCOB05 &  20.82 $\pm$  0.54 &   5.88 $\pm$  0.59 &   0.10 $\pm$  0.01 &   0.99 $\pm$  0.16 &   0.10 $\pm$  0.01 &   3.45 $\pm$  0.94 &   6.61 $\pm$  1.24 &   1.13 $\pm$  0.07 \\
  NCOB06 &  19.74 $\pm$  0.64 &   5.09 $\pm$  0.51 &   0.10 $\pm$  0.01 &   1.09 $\pm$  0.17 &   0.15 $\pm$  0.01 &   4.44 $\pm$  0.93 &   6.61 $\pm$  1.24 &   2.06 $\pm$  0.06 \\
  NCOB07 &  20.37 $\pm$  0.42 &   4.75 $\pm$  0.47 &   0.10 $\pm$  0.01 &   0.94 $\pm$  0.16 &   0.08 $\pm$  0.01 &   4.89 $\pm$  0.87 &   6.61 $\pm$  1.24 &   1.80 $\pm$  0.14 \\
  NCOB08 &  24.25 $\pm$  0.61 &   8.10 $\pm$  0.81 &   0.10 $\pm$  0.01 &   2.13 $\pm$  0.23 &   0.17 $\pm$  0.01 &   4.40 $\pm$  1.03 &   6.61 $\pm$  1.24 &   1.51 $\pm$  0.07 \\
  NCOB09 &  19.92 $\pm$  0.71 &   6.31 $\pm$  0.63 &   0.10 $\pm$  0.01 &   1.02 $\pm$  0.16 &   0.13 $\pm$  0.01 &   4.16 $\pm$  0.93 &   6.61 $\pm$  1.24 &   2.66 $\pm$  0.08 \\
  NCOB10 &  20.53 $\pm$  0.58 &   6.68 $\pm$  0.67 &   0.10 $\pm$  0.01 &   0.93 $\pm$  0.16 &   0.08 $\pm$  0.01 &   3.30 $\pm$  0.81 &   6.61 $\pm$  1.24 &   2.78 $\pm$  0.10 \\
  NCOB11 &  19.28 $\pm$  0.41 &   6.46 $\pm$  0.65 &   0.10 $\pm$  0.01 &   0.95 $\pm$  0.16 &   0.09 $\pm$  0.01 &   3.74 $\pm$  1.00 &   6.61 $\pm$  1.24 &   0.73 $\pm$  0.06 \\
  NCOB12 &  27.99 $\pm$  0.52 &   6.02 $\pm$  0.60 &   0.10 $\pm$  0.01 &   1.64 $\pm$  0.21 &   0.10 $\pm$  0.01 &   9.72 $\pm$  0.92 &   6.61 $\pm$  1.24 &   1.53 $\pm$  0.04 \\
  NCOB13 &  26.74 $\pm$  0.67 &   6.20 $\pm$  0.62 &   0.10 $\pm$  0.01 &   1.76 $\pm$  0.21 &   0.11 $\pm$  0.01 &   8.36 $\pm$  0.99 &   6.61 $\pm$  1.24 &   1.60 $\pm$  0.05 \\
  NCOB14 &  34.86 $\pm$  0.58 &   9.30 $\pm$  0.93 &   0.10 $\pm$  0.01 &   1.83 $\pm$  0.22 &   0.23 $\pm$  0.01 &  13.75 $\pm$  1.11 &   6.61 $\pm$  1.24 &   2.51 $\pm$  0.06 \\
  NCOB15 &  26.73 $\pm$  0.66 &   8.26 $\pm$  0.83 &   0.10 $\pm$  0.01 &   2.18 $\pm$  0.24 &   0.11 $\pm$  0.01 &   5.31 $\pm$  1.06 &   6.61 $\pm$  1.24 &   1.39 $\pm$  0.03 \\
  DCAL01 &  24.01 $\pm$  0.71 &   6.81 $\pm$  0.68 &   0.10 $\pm$  0.01 &   1.04 $\pm$  0.17 &   0.12 $\pm$  0.01 &   6.96 $\pm$  0.89 &   6.61 $\pm$  1.24 &   2.81 $\pm$  0.07 \\
  DCAL02 &  22.88 $\pm$  0.60 &   6.48 $\pm$  0.65 &   0.10 $\pm$  0.01 &   1.05 $\pm$  0.17 &   0.23 $\pm$  0.01 &   7.75 $\pm$  0.91 &   6.61 $\pm$  1.24 &   1.15 $\pm$  0.03 \\
  DCAL03 &  26.02 $\pm$  0.84 &   8.24 $\pm$  0.82 &   0.10 $\pm$  0.01 &   1.91 $\pm$  0.22 &   0.18 $\pm$  0.01 &   8.81 $\pm$  1.15 &   6.61 $\pm$  1.24 &   1.82 $\pm$  0.06 \\
  DCAL04 &  29.36 $\pm$  0.53 &   8.71 $\pm$  0.87 &   0.10 $\pm$  0.01 &   2.04 $\pm$  0.23 &   0.15 $\pm$  0.01 &  13.04 $\pm$  1.15 &   6.61 $\pm$  1.24 &   1.85 $\pm$  0.48 \\
  DCAL05 &  30.23 $\pm$  0.78 &   8.06 $\pm$  0.81 &   0.10 $\pm$  0.01 &   1.90 $\pm$  0.23 &   0.17 $\pm$  0.01 &  10.61 $\pm$  1.29 &   6.61 $\pm$  1.24 &   2.08 $\pm$  0.09 \\
  DCAL06 &  37.86 $\pm$  0.81 &   8.64 $\pm$  0.86 &   0.10 $\pm$  0.01 &   1.64 $\pm$  0.21 &   0.21 $\pm$  0.01 &  19.71 $\pm$  1.16 &   6.61 $\pm$  1.24 &   3.07 $\pm$  0.07 \\
  DCAL07 &  43.98 $\pm$  0.75 &   6.70 $\pm$  0.67 &   0.10 $\pm$  0.01 &   1.81 $\pm$  0.22 &   0.20 $\pm$  0.01 &  24.09 $\pm$  1.24 &   6.61 $\pm$  1.24 &   2.49 $\pm$  0.08 \\
  DCAL08 &  45.42 $\pm$  0.93 &   7.82 $\pm$  0.78 &   0.10 $\pm$  0.01 &   2.20 $\pm$  0.24 &   0.20 $\pm$  0.01 &  24.22 $\pm$  1.31 &   6.61 $\pm$  1.24 &   1.35 $\pm$  0.06 \\
\enddata
\tablecomments{Values in this table are given in units of $\rm nW~m^{-2}~sr^{-1}$. The integrated intensities from faint stars (column 5) and faint galaxies (column 8) cover the apparent magnitude range $19.9 < V \leq 30$ mag AB. The integrated intensities from bright galaxies (column 9) cover the range $V \leq 19.9$ and are derived from DESI/DECam Legacy Survey observations. H$\alpha$ intensities in this table (column 6) are computed from the values (in Rayleigh units) from Table~\ref{tab:irpars} multiplied by the conversion factor 0.24 $\rm nW~m^{-2}~sr^{-1}~R^{-1}$ at 6563\AA. {\mbf The DGL values are computed using eq.~\ref{eqn:dgl_predictor} and the DGL errors are obtained from the Monte Carlo simulations.}}
\end{deluxetable}\label{tab:skyfluxes}

\begin{deluxetable}{lccccccccccccccccc}
\tabletypesize{\scriptsize}
\tablecolumns{17}
\tablewidth{0pt}
\tablecaption{\mbf Systematic and Random Errors for Each Intensity Component}
\tablehead{
\multicolumn{17}{c}{~} \\
\multicolumn{3}{c}{~} &
\multicolumn{4}{c}{Scattered Light from} &
\multicolumn{2}{c}{Faint} &
\multicolumn{2}{c}{H$\alpha$} &
\multicolumn{2}{c}{Diffuse} &
\multicolumn{2}{c}{Faint} &
\multicolumn{2}{c}{Bright}\\
\multicolumn{1}{c}{Field} &
\multicolumn{2}{c}{Total Sky} &
\multicolumn{2}{c}{Stars} &
\multicolumn{2}{c}{Galaxies} &
\multicolumn{2}{c}{Stars} &
\multicolumn{2}{c}{Intensity} &
\multicolumn{2}{c}{Gal. Light} &
\multicolumn{2}{c}{Galaxies} &
\multicolumn{2}{c}{Galaxies} \\
\multicolumn{1}{c}{ID} &
\multicolumn{1}{c}{(sys)} &
\multicolumn{1}{c}{(ran)} &
\multicolumn{1}{c}{(sys)} &
\multicolumn{1}{c}{(ran)} &
\multicolumn{1}{c}{(sys)} &
\multicolumn{1}{c}{(ran)} &
\multicolumn{1}{c}{(sys)} &
\multicolumn{1}{c}{(ran)} &
\multicolumn{1}{c}{(sys)} &
\multicolumn{1}{c}{(ran)} &
\multicolumn{1}{c}{(sys)} &
\multicolumn{1}{c}{(ran)} &
\multicolumn{1}{c}{(sys)} &
\multicolumn{1}{c}{(ran)} &
\multicolumn{1}{c}{(sys)} &
\multicolumn{1}{c}{(ran)} 
}
\startdata
  NHTF01 &    0.16 &  0.75 &    0.62 &  0.00 &    0.01 &  0.00 &    0.17 &  0.06 &    0.00 &  0.01 &    0.94 &  0.13 &    0.97 &  0.78 &    0.11 &  0.03 \\
  NCOB01 &    0.16 &  0.47 &    0.66 &  0.00 &    0.01 &  0.00 &    0.20 &  0.07 &    0.00 &  0.01 &    1.03 &  0.14 &    0.97 &  0.78 &    0.07 &  0.00 \\
  NCOB02 &    0.16 &  0.52 &    0.68 &  0.00 &    0.01 &  0.00 &    0.20 &  0.07 &    0.00 &  0.01 &    1.13 &  0.15 &    0.97 &  0.78 &    0.16 &  0.01 \\
  NCOB03 &    0.16 &  0.61 &    0.66 &  0.00 &    0.01 &  0.00 &    0.20 &  0.07 &    0.00 &  0.01 &    1.00 &  0.14 &    0.97 &  0.78 &    0.06 &  0.01 \\
  NCOB04 &    0.16 &  0.47 &    0.73 &  0.00 &    0.01 &  0.00 &    0.17 &  0.06 &    0.00 &  0.01 &    0.92 &  0.13 &    0.97 &  0.78 &    0.02 &  0.01 \\
  NCOB05 &    0.16 &  0.52 &    0.59 &  0.00 &    0.01 &  0.00 &    0.15 &  0.05 &    0.00 &  0.01 &    0.94 &  0.13 &    0.97 &  0.78 &    0.07 &  0.01 \\
  NCOB06 &    0.16 &  0.62 &    0.51 &  0.00 &    0.01 &  0.00 &    0.16 &  0.06 &    0.00 &  0.01 &    0.92 &  0.13 &    0.97 &  0.78 &    0.06 &  0.01 \\
  NCOB07 &    0.16 &  0.39 &    0.47 &  0.00 &    0.01 &  0.00 &    0.15 &  0.05 &    0.00 &  0.01 &    0.86 &  0.12 &    0.97 &  0.78 &    0.14 &  0.01 \\
  NCOB08 &    0.16 &  0.59 &    0.81 &  0.00 &    0.01 &  0.00 &    0.22 &  0.08 &    0.00 &  0.01 &    1.02 &  0.14 &    0.97 &  0.78 &    0.07 &  0.01 \\
  NCOB09 &    0.16 &  0.69 &    0.63 &  0.00 &    0.01 &  0.00 &    0.15 &  0.05 &    0.00 &  0.01 &    0.92 &  0.13 &    0.97 &  0.78 &    0.08 &  0.02 \\
  NCOB10 &    0.16 &  0.56 &    0.67 &  0.00 &    0.01 &  0.00 &    0.15 &  0.05 &    0.00 &  0.01 &    0.80 &  0.11 &    0.97 &  0.78 &    0.10 &  0.01 \\
  NCOB11 &    0.16 &  0.38 &    0.65 &  0.00 &    0.01 &  0.00 &    0.15 &  0.05 &    0.00 &  0.01 &    0.99 &  0.14 &    0.97 &  0.78 &    0.06 &  0.01 \\
  NCOB12 &    0.16 &  0.49 &    0.60 &  0.00 &    0.01 &  0.00 &    0.20 &  0.07 &    0.00 &  0.01 &    0.91 &  0.12 &    0.97 &  0.78 &    0.04 &  0.02 \\
  NCOB13 &    0.16 &  0.65 &    0.62 &  0.00 &    0.01 &  0.00 &    0.20 &  0.07 &    0.00 &  0.01 &    0.98 &  0.14 &    0.97 &  0.78 &    0.05 &  0.01 \\
  NCOB14 &    0.16 &  0.56 &    0.93 &  0.00 &    0.01 &  0.00 &    0.21 &  0.07 &    0.00 &  0.01 &    1.10 &  0.15 &    0.97 &  0.78 &    0.06 &  0.01 \\
  NCOB15 &    0.16 &  0.64 &    0.83 &  0.00 &    0.01 &  0.00 &    0.23 &  0.08 &    0.00 &  0.01 &    1.05 &  0.14 &    0.97 &  0.78 &    0.03 &  0.02 \\
  DCAL01 &    0.16 &  0.69 &    0.68 &  0.00 &    0.01 &  0.00 &    0.16 &  0.06 &    0.00 &  0.01 &    0.88 &  0.12 &    0.97 &  0.78 &    0.07 &  0.01 \\
  DCAL02 &    0.16 &  0.58 &    0.65 &  0.00 &    0.01 &  0.00 &    0.16 &  0.06 &    0.00 &  0.01 &    0.90 &  0.12 &    0.97 &  0.78 &    0.03 &  0.01 \\
  DCAL03 &    0.16 &  0.82 &    0.82 &  0.00 &    0.01 &  0.00 &    0.21 &  0.07 &    0.00 &  0.01 &    1.14 &  0.16 &    0.97 &  0.78 &    0.05 &  0.03 \\
  DCAL04 &    0.16 &  0.51 &    0.87 &  0.00 &    0.01 &  0.00 &    0.22 &  0.08 &    0.00 &  0.01 &    1.14 &  0.16 &    0.97 &  0.78 &    0.48 &  0.02 \\
  DCAL05 &    0.16 &  0.76 &    0.81 &  0.00 &    0.01 &  0.00 &    0.21 &  0.07 &    0.00 &  0.01 &    1.28 &  0.18 &    0.97 &  0.78 &    0.09 &  0.02 \\
  DCAL06 &    0.16 &  0.79 &    0.86 &  0.00 &    0.01 &  0.00 &    0.20 &  0.07 &    0.00 &  0.01 &    1.15 &  0.16 &    0.97 &  0.78 &    0.07 &  0.02 \\
  DCAL07 &    0.16 &  0.73 &    0.67 &  0.00 &    0.01 &  0.00 &    0.20 &  0.07 &    0.00 &  0.01 &    1.22 &  0.17 &    0.97 &  0.78 &    0.07 &  0.02 \\
  DCAL08 &    0.16 &  0.92 &    0.78 &  0.00 &    0.01 &  0.00 &    0.22 &  0.08 &    0.00 &  0.01 &    1.29 &  0.18 &    0.97 &  0.78 &    0.06 &  0.02 \\
\enddata
\tablecomments{\mbf Error values in this table are given in units of $\rm nW~m^{-2}~sr^{-1}$. The contribution of each error to the final COB and $\rm S_U$ measurements are determined using a Monte Carlo code that properly accounts for the random and systematic errors in every parameter. The ran/sys DGL errors are obtained directly from the Monte Carlo simulations.}
\end{deluxetable}\label{tab:errorbudget}

\begin{figure}[hbtp]
\centering
\includegraphics[keepaspectratio,width=7.0 in]{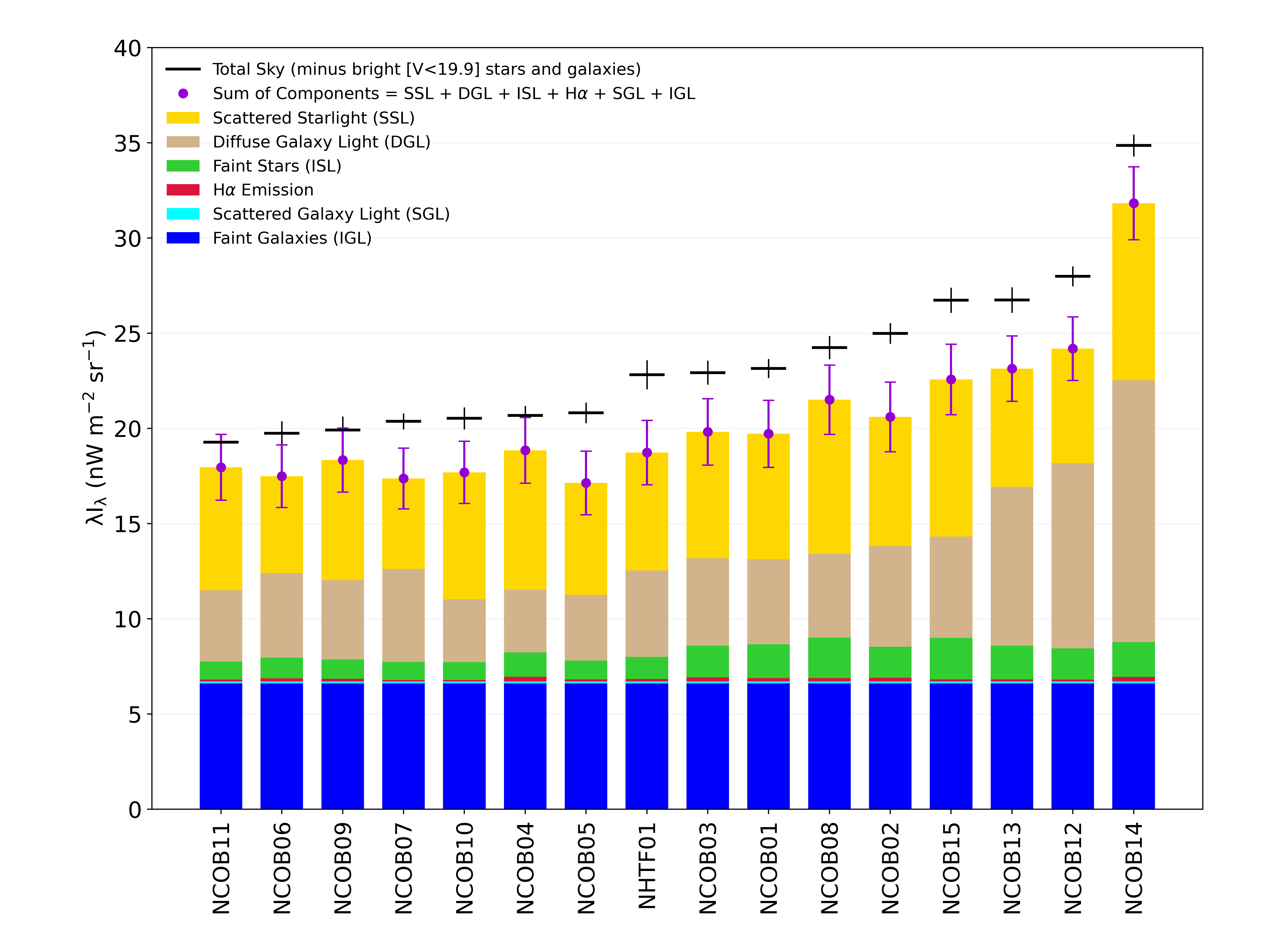}
\caption{A stacked bar chart showing the amplitudes of the known sky components for the 16 COB fields. The black horizontal lines with error bars show the measured total sky values for each ﬁeld, sorted in order of increasing total sky level. The gap between the tops of the bars and the total sky measures represents the anomalous sky intensity in each field. Note that seven fields have total skies of only ${\rm \sim20 ~nW ~m^{-2} ~sr^{-1}}.$ The intensity of bright galaxies in the fields does not contribute to the total skies as measured, but is included in the final COB intensity. {\mbf The purple points are the sum of the known intensity components represented by the stacked bars. The associated errors are provided to show the approximate significance of the gap between the sum of all known components and the total measured sky intensity. The significance of the final COB value is performed using a comprehensive Monte Carlo analysis.}}
\label{fig:bars}
\end{figure}

To visualize the relative importance of all sky components we represent the results as a stacked bar chart for each field in Figure~\ref{fig:bars}. While the summed intensity from every field is less than the total sky level, the dominance of systematic errors in most of the intensity components means the uncertainty in the combined dataset will not be diminished by the $1/\sqrt{N}$ factor. We present our approach for determining the proper propagation of errors for the combined suite of data in the next section.

\begin{figure}[hbtp]
\centering
\includegraphics[keepaspectratio,width=6.0 in]{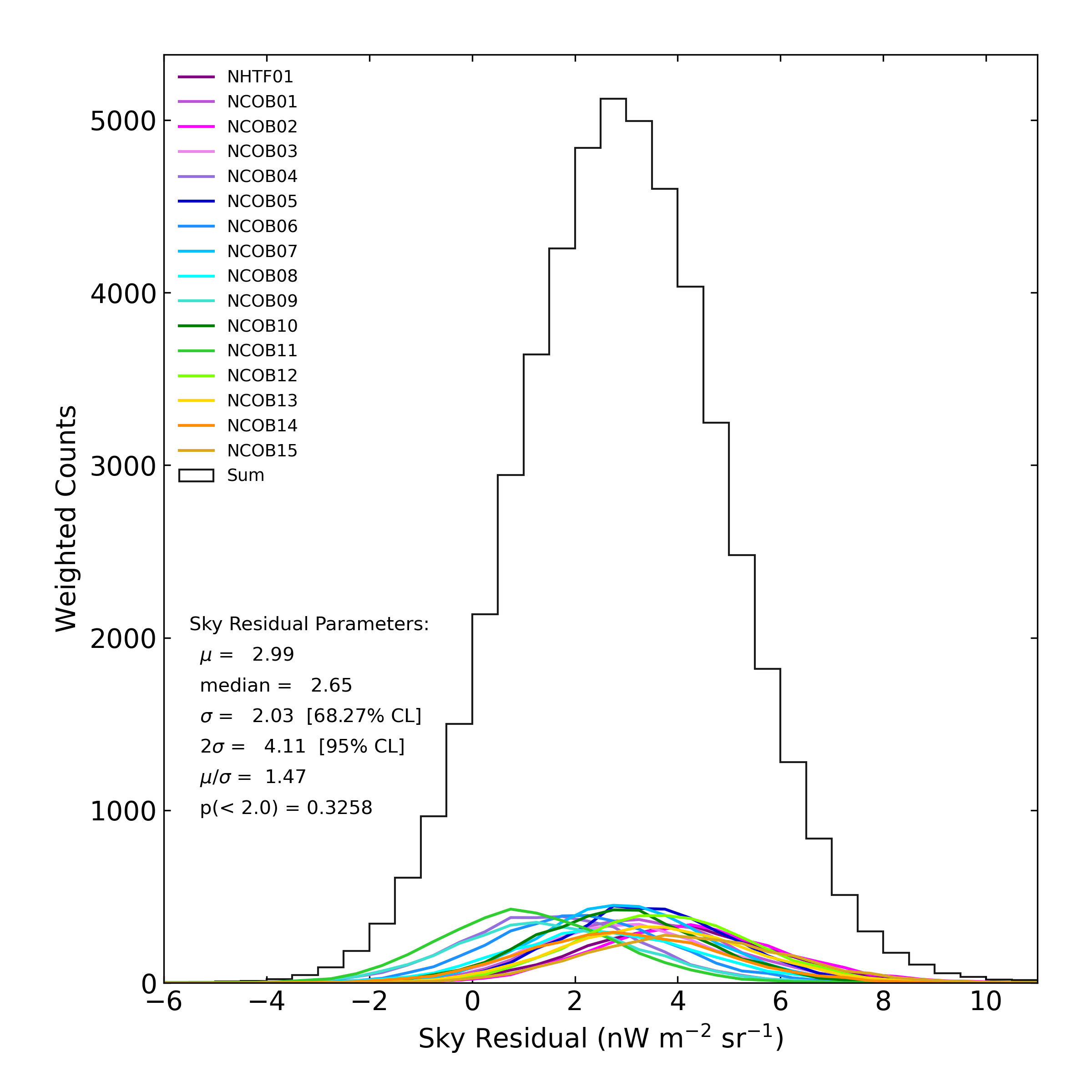}
\caption{The distributions of sky residuals (anomalous intensity) for each of the 16 NCOB fields, produced from 10,000 Monte Carlo realizations of each dataset, are shown as colored curves. The weighted summed distribution for all 16 fields combined is shown by the black histogram. The sky residual is the intensity left over after all known sources of light are subtracted from the measured total sky level. If all known sources of light had been accounted for, then the peak of the summed distribution would be centered on zero. The actual distribution is shifted from zero but only by {\mbf +1.47}$\sigma$.}
\label{fig:dcob_mc}
\end{figure}

\begin{figure}[hbtp]
\centering
\includegraphics[keepaspectratio,width=6.0 in]{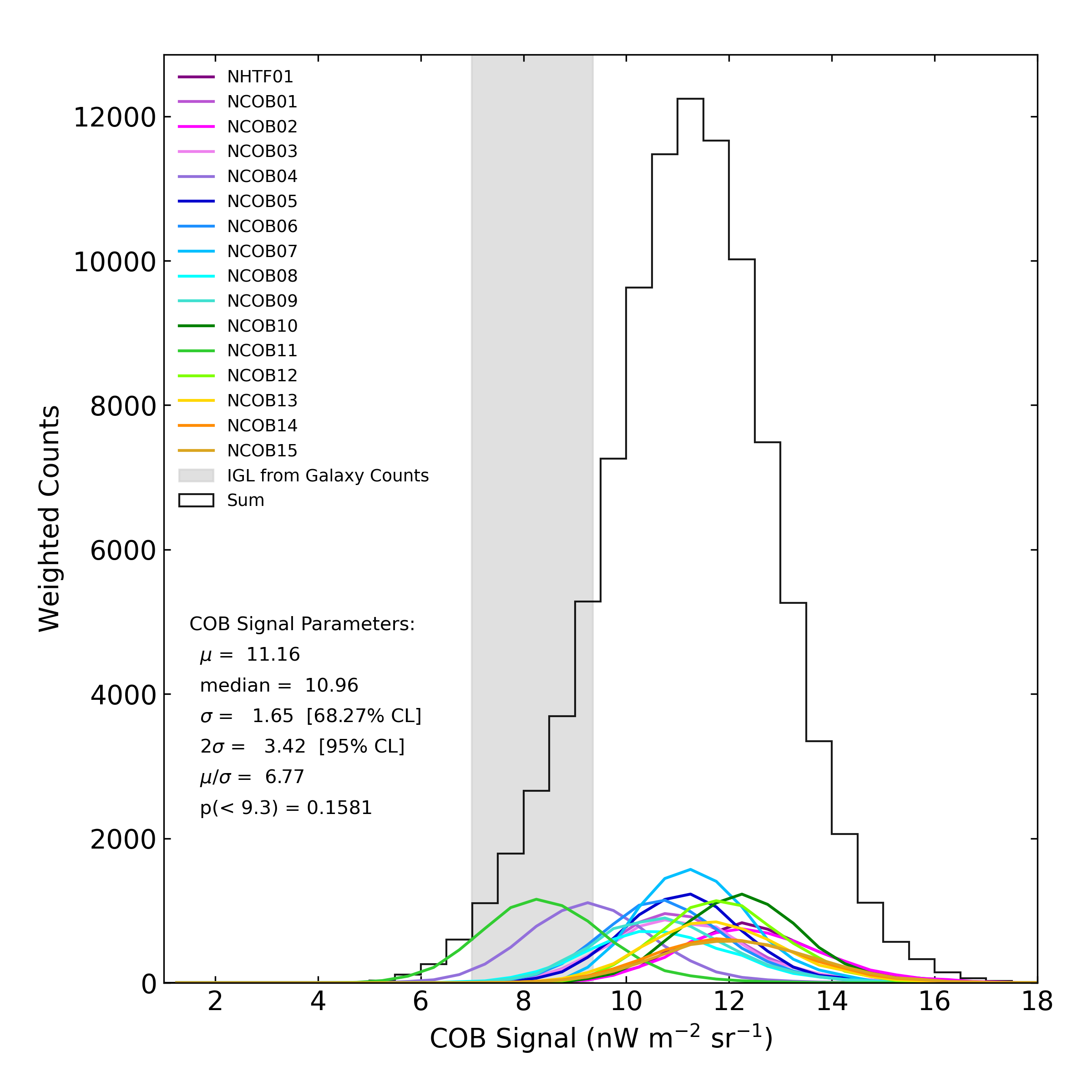}
\caption{The distributions of Cosmic Optical Background (COB) values for each of the 16 NCOB fields, produced from 10,000 Monte Carlo realizations of each dataset, are shown as colored curves. The weighted summed distribution for all 16 fields combined is shown by the black histogram. The predicted extragalactic background intensity in the optical passband from a variety of galaxy surveys is shown by the vertical gray bar. The width of the gray bar is set by the uncertainty in the integrated light from known galaxies. The mean of the measured COB distribution from this work differs from zero by {\mbf +6.77}$\sigma$, representing one of the most significant detections of the COB to date. }
\label{fig:cob_mc}
\end{figure}

\section{The Cosmic Optical Background}\label{sec:cob}

\subsection{A Monte Carlo Approach to Estimating the COB Intensity}

{\tbf Having identified all the sources of foreground optical emission known to us in the two previous sections, the task is then to recover an optimal estimate of the COB intensity with accurate errors that reflect the appropriate random and systematic uncertainties (see Table~\ref{tab:errorbudget}), as well as any covariances  among the parameters used to estimate the COB intensity.  We will use a Monte Carlo approach that randomly generates complete realizations of the COB observations as based on our error model.  As noted at the start of $\S\ref{sec:decomp},$ we do this in two steps. With the exception of the DGL component, our knowledge of all other foregrounds, as well as the total sky level itself, comes from mutually independent information.  As such, we start with equation (\ref{eqn:IDGL}) to estimate DGL+, the DGL intensity plus any anomalous intensity component, which is not affected by covariance among the terms that are subtracted off to isolate it. The second step is to then use the FIR background intensity to isolate the DGL component itself.  As this uses the observations to develop a self-calibrated DGL estimator, this step does account for covariance between the observational parameters.}

{\tbf For the first step then, the Monte Carlo routine generates 10,000 realizations of the total sky intensity observations and non-DGL foreground components ($S_T,~IGL,~SSL,~SGL,~ISL,~BIGL,~{\rm and}~I(H\alpha)$) for each field. We also generate random realizations of the {\it independent} variables (FIR intensities and cosmic IR backgrounds) that will be used to estimate the DGL.  In detail, for any observed component or observational input, $D_{obs}(j),$ in field $1\leq j, \leq24$ (16 NCOB $+$ 8 DCAL fields), the Monte Carlo routine generates a set of simulated values for $1\leq i\leq10^4:$
\begin{equation}\label{eqn:dsim}
    \rm D_{sim}(i,j) = D_{obs}(j) ~+~ \sigma_{RAN}(j) G_{D}(i,j) ~+~ \sigma_{SYS}(j) F_D(i),
\end{equation}
where, $G_{D}(i,j)$ and $F_D(i)$ are sequences of gaussian random variables with zero mean and unit variance.  Note that $F_D(i)$ scales the systematic error term $\sigma_{SYS}(j)$ in any field, and is thus the same for all fields for a given $i$ realization. For many parameters, $\sigma_{SYS}$ will also be constant over all fields, but relative systematic errors may vary from field to field in step with the given parameter (as occurs in the case of SSL, for example).  In contrast, $G_{D}(i,j)$ scales the random errors, which vary freely from field to field in any given realization.  Lastly, we note that while equation (\ref{eqn:dsim}) is fully general in allowing for both systematic and random errors for any component, in practice the systematic errors strongly dominate the total error budget, and for many components the random error term is negligible.} 

\subsection{Estimating the DGL Intensities} 

{\tbf A two-band DGL estimator of the form given by equation (\ref{eqn:2-band}) is generated for every Monte Carlo realization from its set of 24 DGL+ values produced, which is used to predict the DGL backgrounds {\it for that particular realization.}  As is described by equation (\ref{eqn:cibsubflux}), the independent variables used by the estimator are the FIR background intensities at 350$~\mu$m and 550$~\mu$m, with the estimated CIB backgrounds subtracted. Fitting the DGL+ values to the FIR background intensities generates the $c_1,$ $c_2,$ and $c_3$ coefficients in equation (\ref{eqn:2-band}).  The best-fit DGL-alone estimates for each field are then provided by the $c_2,$ and $c_3$ coefficients on the assumption that DGL is zero for zero FIR background intensity.  This directly highlights the strong covariance between the assumed CIB backgrounds and the DGL predictions noted in $\S\ref{sec:FIRCIB}.$  This covariance is included in the Monte Carlo routine, however, simply by including errors in the all the FIR parameters as part of the analysis. The coefficients of the DGL estimator are also covariant, but again by having the Monte Carlo routine sample the error distributions of both the independent and dependent variables used by the DGL estimator, this behavior is included implicitly in the analysis as well.  Further, because the DGL+ intensities for any field reflect the errors of all components used to generate them, the Monte Carlo analysis also incorporates the full suite of covariances associated with deriving the DGL estimator from the COB observations, themselves. 

Lastly, we note that while we presented a {\it mean} two-band DGL estimator in equation (\ref{eqn:dgl_predictor}), we actually do not use it as such for the final analysis.  Again, we generate the DGL estimator anew for each Monte Carlo realization.  Our interest is really just in the distribution of DGL estimates for any given field generated by the full ensemble of Monte Carlo runs.  These distributions, in fact, rather than any simple error calculation, provide the appropriate uncertainties in the DGL values needed to derive the errors in the final COB and $S_U$ intensities.}

\pagebreak
\subsection{Estimating the COB Intensity and Anomalous Background Component} 

{\tbf With the DGL component derived for all fields in all 10,000 Monte Carlo runs, it is simple to derive the COB intensity and any residual anomalous intensity background, ${\rm S_U,}$ for each realization. The optimal estimate of both parameters for the whole survey will be derived by first generating the COB and ${\rm S_U}$ intensity distributions for each field, then combining them into a full ensemble distribution weighted by the inverse variance of the individual-field distributions.}

The anomalous component, ${\rm S_U,}$ is just the residual intensity remaining after all known sources of light are subtracted from the data. Specifically,
\begin{equation}\label{eqn:resid_sky}
    \rm S_U = S_T - IGL - SSL - SGL - ISL - I(H\alpha) - DGL.
\end{equation}
We compute $\rm S_U$ for each field using the 10,000 realizations of each parameter on the right hand side of equation~\ref{eqn:resid_sky}. We then generate the cumulative distribution function (CDF) of $\rm S_U,$ for each field and find its 68.3\% confidence limits. We adopt the 1-$\sigma$ value on $\rm S_U$ to be half the difference between high and low 68.3\% confidence limits for that specific field. We then generate a summed distribution of all the $\rm S_U$ values for all fields, weighting the $\rm S_U$ values for each field by the inverse square of that field's corresponding 1-$\sigma$ value. The value of $\rm S_U$ for the full sample is then taken to be the mean value of that weighted sum distribution, and the error in the full sample $\rm S_U$ is taken as the half width of the 68.3\% confidence limit range of that weighted sum distribution. The resulting summed distribution of $\rm S_U$ values for all 16 COB survey fields is shown in Figure~\ref{fig:dcob_mc}. 

To get the {\tbf optimal full-survey} COB intensity, we repeat the above procedure but apply it to the following expression:
\begin{equation}\label{eqn:cob_sky}
\begin{split}
     \rm COB &= \rm S_T - SSL - SGL - ISL - I(H\alpha) - DGL + BIGL \\
             &= \rm S_U + IGL + BIGL.
\end{split}
\end{equation}
We remind the reader that we need to add back the integrated light of bright galaxies ($\rm BIGL$) prior to computing the COB since all galaxies brighter than $V=19.9$ had been masked out prior to measuring the total sky level (see $\S\ref{sec:igl}$ for details). The resulting summed distribution of COB values for all 16 survey fields is shown in Figure~\ref{fig:cob_mc}. 

Our derived COB and $\rm S_U$ values are:
\begin{equation}\label{eqn:final_results}
    \begin{split}
        \rm COB & {\mbf = 11.16 \pm 1.65 ~(\rm 1.47 ~sys, ~0.75 ~ran)} \rm ~nW~m^{-2}~sr^{-1} \\
        \rm S_U & {\mbf = \phantom{1}2.99 \pm 2.03 ~(\rm 1.75 ~sys, ~1.03 ~ran)} \rm ~nW~m^{-2}~sr^{-1} 
    \end{split}
\end{equation}
The COB is detected at a significance of {\mbf 6.77}$\sigma$ but the anomalous (residual) sky intensity is only present at the {\mbf 1.47}$\sigma$ level. In other words, $\rm S_U$ is likely consistent with zero.

\subsection{Jackknife Tests of the Robustness of the Results}

\begin{figure}[hbtp]
\centering
\includegraphics[keepaspectratio,width=6.0 in]{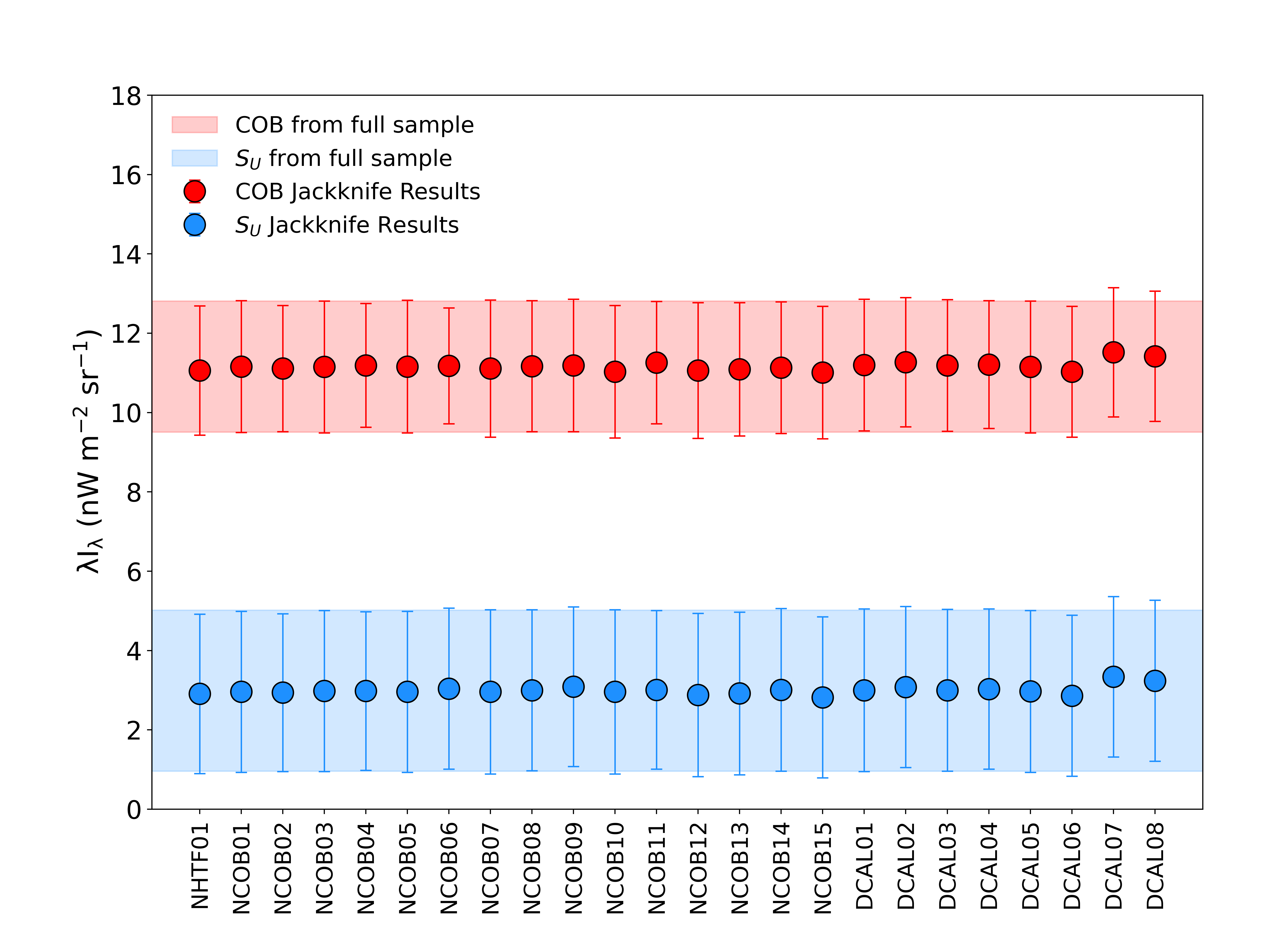}
\caption{Jackknife test results are shown for the COB and ${\rm S_U}$ intensity measurements. The data points show the intensities calculated when the indicated NCOB or DCAL field (shown along x-axis) is excluded from the analysis. The center y-values and heights of the red and blue horizontal bands are the derived values and their 1-$\sigma$ ranges for the full-sample COB and $\rm S_U$, respectively. No field is seen to have an outsized influence on the results using the total sample.}
\label{fig:jackknife}
\end{figure}

To determine if any one of our NCOB or DCAL fields has an outsized influence on either the COB or $\rm S_U$ estimate, we perform a series of jackknife tests, where for each test we exclude just one field from the Monte Carlo simulations, doing this in turn for each field in the full set of 16 NCOB and 8 DCAL fields. The outcomes of these tests are shown in Figure~\ref{fig:jackknife}. The x-axis labels show the name of the field that was excluded in each 10,000 run simulation. The red and blue data points show, respectively, the resulting COB and $\rm S_U$ value derived when that field was excluded from the analysis. The standard deviation in the COB or $\rm S_U$ for these 24 jackknife tests is just $0.12 \rm ~nW~m^{-2}~sr^{-1}$. We conclude that the COB and $\rm S_U$ results are resilient against the removal of any one of the NCOB or DCAL fields from the analysis. 

\subsection{Interplanetary Dust (IPD) Field Results}

As noted in $\S\ref{sec:design},$ we obtained observations of four fields at low ecliptic latitude to verify the assumption that there is no significant zodiacal light emission at heliocentric distances beyond 50 AU. We perform this test by computing the COB and $\rm S_U$ intensity values for the four IPD fields and look for any significant deviations from the main survey results listed in equation~\ref{eqn:final_results}. The results are given in Table~\ref{tab:ipdf_results}. Because the IPDF observations were not part of our rigorous survey selection process, the errors on the COB and $\rm S_U$ values for the IPD fields are estimated analytically rather than with the Monte Carlo procedure but should be approximately similar to what the MC process would yield. Columns 3 and 5 in Table~\ref{tab:ipdf_results} give the difference between the COB and $\rm S_U$ given in equation~\ref{eqn:final_results} and the corresponding values for each IPD field, respectively. We find no significant differences between the COB and $\rm S_U$ values obtained for the IPD Fields and those derived from the COB science fields supporting the assertion that zodiacal light is negligible at the distances where NH made the COB observations presented here. The consistency between the COB science observations and the IPDF observations also suggests that there are no major zodi-subtraction residuals in the Planck HFI maps.

\begin{deluxetable}{ccccc}
\tabletypesize{\small}
\tablecolumns{5}
\tablewidth{0pt}
\tablecaption{COB and $\rm S_U$ Results for the IPD Fields}
\tablehead{
\multicolumn{5}{c}{~}\\
\multicolumn{1}{c}{Field} &
\multicolumn{1}{c}{COB} &
\multicolumn{1}{c}{$\Delta$COB} &
\multicolumn{1}{c}{$\rm S_U$} &
\multicolumn{1}{c}{$\Delta \rm S_U$} \\
\multicolumn{1}{c}{ID} &
\multicolumn{1}{c}{Value} &
\multicolumn{1}{c}{(NCOB-IPDF)} &
\multicolumn{1}{c}{Value} &
\multicolumn{1}{c}{(NCOB-IPDF)} 
}
\startdata
IPDF01 & \phantom{1}8.78 $\pm$ 2.59 &  $\phantom{-}2.38 \pm 3.07$  &  0.61 $\pm$ 2.88  & $\phantom{-}2.38 \pm 3.52$  \\
IPDF02 &           11.24 $\pm$ 2.79 &  $-0.08 \pm 3.24$  &  3.18 $\pm$ 3.05  & $-0.19 \pm 3.66$  \\
IPDF03 &           12.10 $\pm$ 2.14 &  $-0.94 \pm 2.70$  &  2.69 $\pm$ 2.47  & $\phantom{-}0.30 \pm 3.20$  \\
IPDF04 &           11.15 $\pm$ 2.02 &  $\phantom{-}0.01 \pm 2.61$  &  3.44 $\pm$ 2.36  & $-0.45 \pm 3.11$  \\
 \enddata
\tablecomments{All values are in units of $\rm nW~m^{-2}~sr^{-1}$.}
\end{deluxetable}\label{tab:ipdf_results}

\begin{figure}[hbtp]
\centering
\includegraphics[keepaspectratio,width=6.0 in]{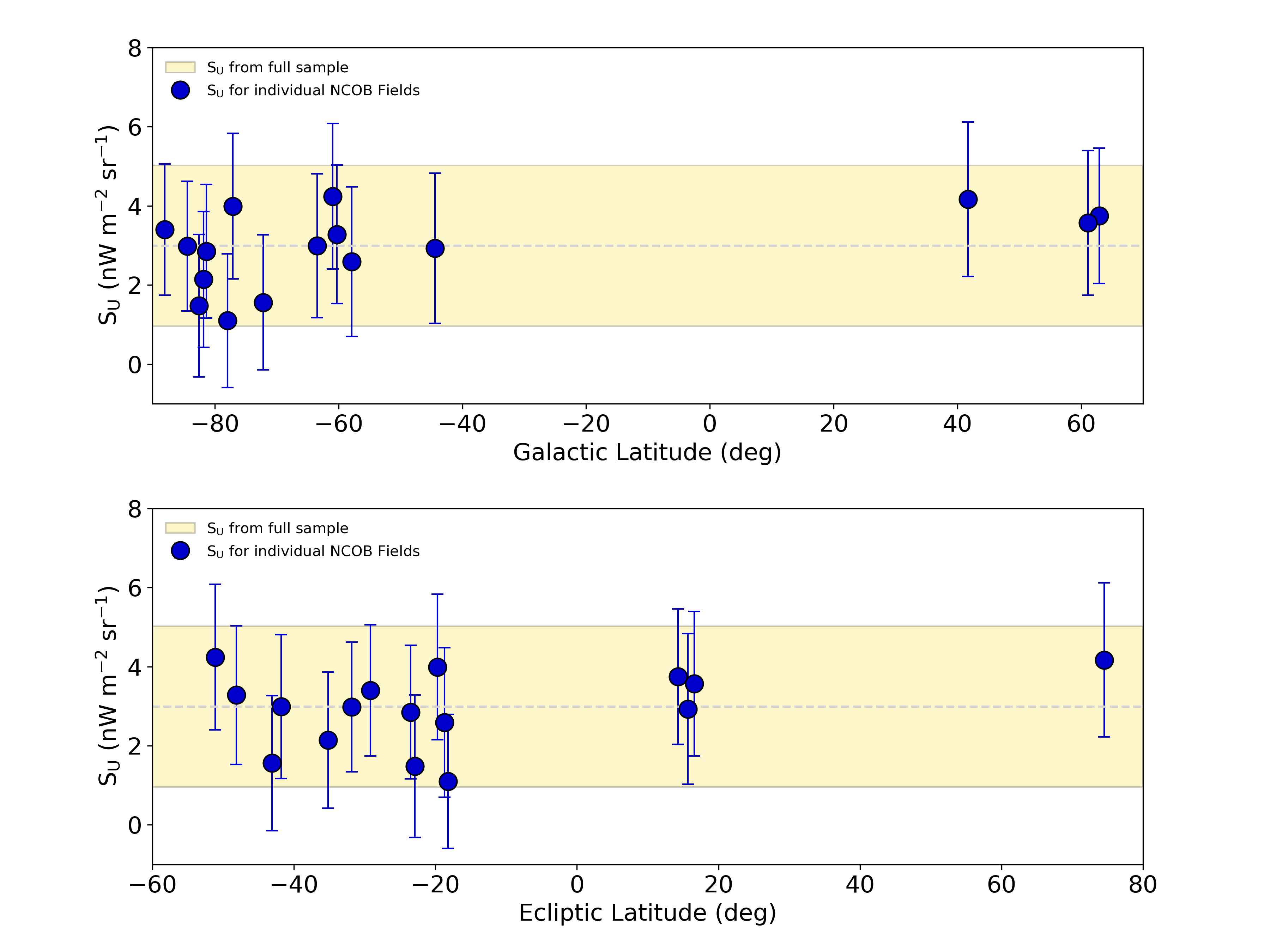}
\caption{{\mbf The sky residual signal, $\rm S_U$, as a function of galactic latitude (top) and ecliptic latitude (bottom) for each individual NCOB field. The yellow horizontal band shows the sky residual $\pm 1$ sigma range for the full sample of 16 fields. No significant trend is seen between $\rm S_U$ and ecliptic or galactic latitude.}}
\label{fig:su_vs_lat}
\end{figure}
\pagebreak
\subsection{Isotropy of the COB}

With 16 fields spread over the sky, we can, in principle, look for anisotropy in the COB. Of the 16 NCOB fields, 13 are in the south Galactic hemisphere (SGH) and 3 are in the north Galactic hemisphere (NGH). We derive the COB for these two subsets to see if there are any significant offsets between them. We use the 10,000 realizations generated and separate out the north and south Galactic subsamples from them. The COB and $\rm S_U$ values are then calculated. The results are shown in Table~\ref{tab:cob_by_hemisphere}. {\mbf We also plot the individual $\rm S_U$ values as a function of ecliptic and galactic latitude in Figure~\ref{fig:su_vs_lat}.} We do not see any significant difference between the COB or $\rm S_U$ values derived for the NGH vs SGH subsamples. {\mbf Nor do we seen any significant trend with ecliptic latitude with the current sample.} We note that with only 3 NCOB fields in the NGH and only 3 DCAL fields in the SGH, we use the full sample to compute the two-band DGL estimator in this isotropy analysis. A larger NGH NCOB sample and more DCAL observations in the SGH would be required to do a more in-depth exploration of any anisotropy in the COB signal.

\begin{deluxetable}{ccrcrc}
\tabletypesize{\small}
\tablecolumns{6}
\tablewidth{0pt}
\tablecaption{North / South Galactic Hemisphere COB and $\rm S_U$ Results}
\tablehead{
\multicolumn{6}{c}{~}\\
\multicolumn{1}{c}{Galactic} &
\multicolumn{1}{c}{COB} &
\multicolumn{1}{c}{Stat.} &
\multicolumn{1}{c}{$\rm S_U$} &
\multicolumn{1}{c}{Stat.} &
\multicolumn{1}{c}{Number of} \\
\multicolumn{1}{c}{Hemisphere} &
\multicolumn{1}{c}{Value} &
\multicolumn{1}{c}{Signif.} &
\multicolumn{1}{c}{Value} &
\multicolumn{1}{c}{Signif.} &
\multicolumn{1}{c}{Fields}
}
\startdata
 North   &  11.91 $\pm$ 1.34 &  8.92  &  4.16 $\pm$ 1.96  & 2.12  &  3 \\
 South   &  10.93 $\pm$ 1.64 &  6.65  &  2.80 $\pm$ 2.00  & 1.40  & 13 \\
 \enddata
\tablecomments{COB and $\rm S_U$ values are in units of $\rm nW~m^{-2}~sr^{-1}$.}
\end{deluxetable}\label{tab:cob_by_hemisphere}

\section{Convergence}

Broadly speaking, there are three ways to measure the intensity of the cosmic optical background. In one approach, determining the COB intensity is a natural outcome of a panoramic and lengthy campaign to identify and assay all light sources in the Universe.  In practice, this means beginning with a complete census of galaxies and their associated active nuclei.  Galaxies exist, therefore there must be a COB. The tally of known light-emitting objects  always defines the lower limit to the COB intensity.  A second approach is inferential.  The existence of the COB implies that very high energy (VHE) $\gamma$-rays cannot freely traverse the Universe.  Their observed extinction as a function of cosmological distance to their source AGNs provides an estimate of the COB intensity. The final third approach is that attempted here: direct observation of the COB intensity. This requires care to isolate and correct for irrelevant foreground intensity sources, but also allows for the discovery of previously unknown intensity sources.

\begin{figure}[hbtp]
\centering
\includegraphics[keepaspectratio,width=7.0 in]{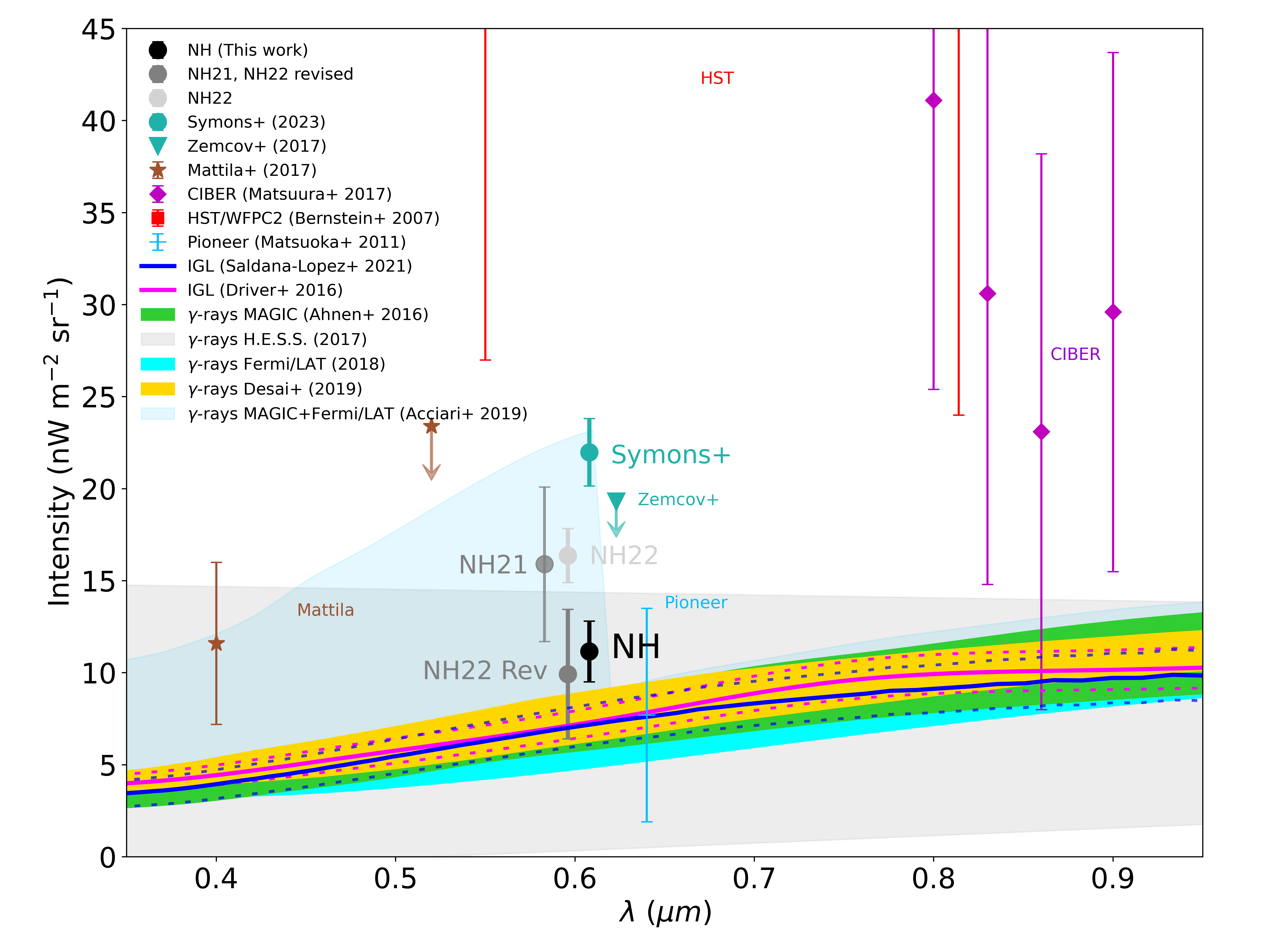}
\caption{The present result is compared to previous COB measures over the wavelengths spanned by the LORRI passband. Direct COB intensity measurements are shown as points with error bars. The NH21 and NH22 intensities are slightly offset to the blue for clarity. The \citet{zemcov} intensity-limit (offset to the red for clarity) and the \citet{mattila} 0.52 ${\rm\mu m}$ limit are shown as 2-$\sigma$ upper limits. The NH22 intensity is shown twice.  The upper point is the intensity as published in NH22.  The lower point is the NH22 intensity revised to correct the DGL subtraction error noted in $\S\ref{sec:dgl}.$ IGL estimates are shown as lines with 1-$\sigma$ bounds. COB intensities inferred from VHE $\gamma$-rays are shown as shaded bands.}
\label{fig:comp}
\end{figure}

At the outset of this work we posed the question: Is the COB intensity as expected from our census of faint galaxies, or does the Universe comprise additional sources of light not yet recognized? With our present result, it appears that these diverse approaches are converging to a common answer.  Galaxies are the greatly dominant and perhaps even complete source of the COB. There does remain some room for interesting qualifications and adjustments to this picture, but in broad outline it is the simplest explanation for what we see.

Figure \ref{fig:comp} shows our present result in the context of COB measurements from all three methods. We presented a previous version of this figure in NH22, but we revisit it here in light of our revised estimate of the COB intensity and the greatly reduced allowance for an anomalous COB component.  As we noted in NH22, there is excellent agreement on the IGL level over the ensemble of estimates.
\citet{driver16}, \citet{saldana}, and our own estimate (see NH21), all imply a contribution to the COB intensity of $\sim 8{\rm  ~nW ~m^{-2} ~sr^{-1}}$ over the passband sampled by LORRI.  At the same time, the galaxy counts feeding into the IGL are obtained from similar, if not the same, observational sources, and thus may have common systematic errors.  For example, \citet{conselice} argued that the galaxy counts are seriously incomplete, while \citet{cooray12}, \cite{zem14}, and \citet{mat19} argued that the COB includes a substantial component of light from stars tidally removed from galaxies, or from a population of faint sources in extended halos. Our present COB intensity would indeed allow for a modest enhancement in the implied starlight contribution to the COB, but not a wholesale revision of it. To explain the $\rm S_U$ value of ${\rm 2.91\pm2.03 ~nW ~m^{-2} ~sr^{-1}}$ as extragalactic in origin would require {\mbf a $\sim37 (\pm 7)$\%} increase in light from galaxies or intergalactic space {\mbf corresponding to the ratio of our COB value ($\rm 11.16 ~nW ~m^{-2} ~sr^{-1}$) to that predicted from deep galaxy counts ($\rm = IGL + ~BIGL = 8.17 ~nW ~m^{-2} ~sr^{-1}$)}. \cite{driver16} suggest a diffuse component to extragalactic background light could be present at the 20\% level, possibly due to low surface brightness galaxies and/or intrahalo light in the specific case of the COB spectrum accessible to LORRI. We note that if we extend the faint end galaxy count integration limit to $V=34$ mag instead of $V=30$ mag and we assume that the faint end slope of the galaxy number counts $-$ magnitude relation remains unchanged, our IGL estimate would increase by $\sim8$\% with a corresponding reduction in $\rm S_U$.

We show the COB constraints from five recent VHE ($0.1-30$ TeV) $\gamma$-ray studies: \citet{magic16},  \citet{hess17}, \citet{fermi}, \citet{fermi19}, and \citet{magic19} in Figure \ref{fig:comp}. The concordance of the COB inferred from galaxy counts and VHE $\gamma$-ray absorption has already long been advanced as a strong argument that the COB is mainly due to the light of known galaxies. A significantly higher COB intensity would engender significantly higher VHE $\gamma$-ray extinction.  One interesting caveat, however, is that most of these studies assume that the spectral energy distribution of the COB photons is the same as that of the integrated galaxy light. When the analysis allows for arbitrary intensity as a function of wavelength, as was done in the \citet{hess17} and \citet{magic19} papers (shown in Figure \ref{fig:comp} with light shading), the VHE $\gamma$-ray constraints can allow for considerably larger COB intensity than that associated with the IGL alone.

Figure \ref{fig:comp} shows several examples of direct-detection COB measures made from near-Earth space that fall within the LORRI passband.  As was discussed in NH22, these include the HST/WFPC2 observations of \citet{wfpc2}, the CIBER rocket-based measures of \citet{ciber}, and the ``dark cloud" measures of \citet{mattila}. The 0.40 ${\rm\mu m}$ intensity of \citet{mattila} and the 0.80 ${\rm\mu m}$ CIBER intensity of \citet{ciber} reject a null detection of the COB with only slightly better than 2-$\sigma$ significance.  Most of the measures are not significantly different than zero.  More recently, the SKYSURF project \citep{skysurf} attempted to detect the COB  in three NIR bands to the red of 1 ${\rm\mu m}$, using archival HST observations, but only achieved upper limits of  29 ${\rm~nW~m^{-2}~sr^{-1}}.$  It is still extremely difficult to get past the strong effects of zodiacal light in the inner solar system.

Our present COB measure of ${\rm 11.16 \pm 1.65  ~nW ~m^{-2} ~sr^{-1}}$ is plotted in Figure \ref{fig:comp} with the ``NH" label. The most important contrast with our earlier work, is the present 32\%\ downward revision of the COB intensity as compared to the ${\rm 16.37\pm 1.47  ~nW ~m^{-2} ~sr^{-1}}$ intensity in NH22, based on NHTF01. As noted in $\S\ref{sec:dgl},$ we concluded that the DGL contribution to NHTF01 had been seriously underestimated due to an incorrect correction that we applied to the foreground 100 ${\rm\mu m}$ intensity that we used with the \citet{zemcov} DGL estimator.
In Figure \ref{fig:comp} we also plot the NH22 result with the revised 100 ${\rm\mu m}$ intensity as ``NH22 Rev" {\tbf to demonstrate the effect of this revision alone.} The error bars increase with the larger implied DGL correction, but this intensity is now in excellent agreement with the present COB intensity.
{\tbf Again, the NHTF01 image set is fully included in our present analysis with all the revisions noted in Table \ref{tab:program}, including the new DGL estimator developed in $\S\ref{sec:dgl}$ and the count-rate decay correction presented in $\S\ref{sec:decay},$}

We note that our present COB intensity is only $\sim50\%$ of the ${\rm 21.98\pm 1.23~(ran) \pm 1.36~(sys)  ~nW ~m^{-2} ~sr^{-1}}$ COB intensity reported by \citet{symons}. As that measure is made with LORRI as well, and indeed incorporates all the archival data presented in NH21 (albeit with additional archival data that we chose not to use in NH21), this is concerning. At present we can not resolve this difference.  We can only remark that the \citet{symons} analysis was independent of ours and made several choices concerning the archival data used and the detailed processing that differ from ours.  We also note that the raw {\it total} sky intensities for seven of our 16 fields before any foreground intensity sources were subtracted are already less than the {\it final} \citet{symons} COB intensity (see Figure \ref{fig:bars} or Table \ref{tab:skyfluxes}).  

If our present COB intensity is correct, however, it means that galaxy counts, VHE $\gamma$-ray extinction, and direct optical band measures of the COB intensity have finally converged at an interesting level of precision. There is still room to adjust the galaxy counts slightly, or to allow for non-dominant anomalous intensity sources.  But the simplest hypothesis appears to provide the best explanation of what we see: the cosmic optical background is the light from all the galaxies within our horizon.

\acknowledgments

We thank NASA for funding and continued support of the New Horizons mission, which were required to obtain the present observations. No New Horizons NASA funds were used, however, for the reduction and analysis of the 2023 NH COB observations.  The data presented were obtained during the second Kuiper Extended Mission of New Horizons. We thank Bruce Draine and Jean-Marc Casandjian for useful conversations. {\mbf We thank the anonymous referee for their careful review and insightful comments that helped us to improve the clarity of this paper.}
MP is funded by the Space Telescope Science Institute, which is operated by the Association of Universities for Research in Astronomy, Inc. (AURA), under NASA contract NAS 5–26555.  TRL is funded by NSF NOIRLab, which is managed by AURA under a cooperative agreement with the National Science Foundation.
We also thank the Aspen Center for Physics for their hospitality during our time completing this paper. The Aspen Center for Physics is supported by National Science Foundation grant PHY-2210452. 

This work made use of data from the European Space Agency (ESA) mission {\it Gaia} (\url{https://www.cosmos.esa.int/gaia}), processed by the {\it Gaia} Data Processing and Analysis Consortium (DPAC, \url{https://www.cosmos.esa.int/web/gaia/dpac/consortium}). Funding for the DPAC has been provided by national institutions, in particular the institutions participating in the {\it Gaia} Multilateral Agreement.
This work also made use of data obtained with Planck (\url{https://www.esa.int/Science_Exploration/Space_Science/Planck}), an ESA science mission with instruments and contributions directly funded by ESA Member States, NASA, and Canada.

This research uses services or data provided by the Astro Data Lab, which is part of the Community Science and Data Center (CSDC) program at NSF's National Optical-Infrared Astronomy Research Laboratory. NOIRLab is operated by
the Association of Universities for Research in Astronomy (AURA), Inc. 
under a cooperative agreement with the National Science Foundation.
This research has also made use of the NASA/IPAC Infrared Science Archive, which is funded by the National Aeronautics and Space Administration and operated by the California Institute of Technology.

\software{astropy \citep{2013A&A...558A..33A, 2018AJ....156..123A}, matplotlib \citep{matplotlib},  TRILEGAL \citep{tril2005}, Vista \citep{vista}}

{}

\end{document}